\documentclass[10pt,eqsecnum,preprint,flushrt]{aastex}
\slugcomment{September 1, 2001}
\newcommand{\be}{\begin{equation}}
\newcommand{\ee}{\end{equation}}

\newcommand{\Et}{\tilde E}
\newcommand{\lt}{\tilde l}

\def\R#1#2{R_{(#1)(#2)}}
\def\T#1#2{T_{(#1)(#2)}}
\def\tet#1#2{e_{(#1)}^{\phantom{(#1)}#2}}
\def\gvvc{\raisebox{-2ex}{$\qquad\gamma\;\;$}{$v/v_c$}}

\begin{document}

\title{RELATIVISTIC SELF-SIMILAR DISKS}

\author{Mike J. Cai}
\affil{Department of Physics, University of California \\
              Berkeley, CA 94720-3411, USA}
\email{mcai@astron.berkeley.edu}
\author{Frank H. Shu}
\affil{Department of Astronomy, University of California \\
              Berkeley, CA 94720-3411, USA}
\email{fshu@astron.berkeley.edu}

\begin{abstract}
We formulate and solve by semi-analytic means the axisymmetric
equilibria of relativistic self-similar disks of infinitesimal
vertical thickness.  These disks are supported in the horizontal
directions against their
self-gravity by a combination of isothermal (two-dimensional)
pressure and a flat rotation curve.  The dragging of inertial
frames restricts possible solutions to rotation speeds
that are always less than 0.438 times the speed of light,
a result first obtained by Lynden-Bell and Pineault in
1978 for a cold disk.  We show that prograde circular orbits
of massive test particles exist and are stable for
all of our model disks, but retrograde circular orbits
cannot be maintained with particle velocities less than
the speed of light once the disk develops an ergoregion.
We also compute photon trajectories,
planar and non-planar, in the resulting spacetime,
for disks with and without ergoregions.  We find that
all photon orbits, except for a set of measure zero,
tend to be focused by the gravity of the flattened mass-energy
distribution toward the plane of the disk.  This result
suggests that strongly relativistic, rapidly rotating, compact objects
may have difficulty ejecting collimated beams of matter or light along
the rotation axes until the flows get well beyond the
flattened parts of the relativistic mass distribution
(which cannot happen in the self-similar models considered
in this paper).

\end{abstract}

\keywords{general relativity: disks; general relativity:
ergoregion; general relativity: frame
dragging; general relativity: light bending}

\section{Introduction}
\label{sec:intro}

It has been said that there are basically
only two kinds of self-gravitating objects in astronomy, spheres and disks.
It has also been said that general relativity is so beautiful, it has to
be right.  Thus, spheres and disks should be of as much natural
interest to relativists as to Newtonian dynamicists.  Yet because the
beauty of general relativity comes at the steep price
of great mathematical difficulty, for many decades the
only known solutions to Einstein's field equations were ones
possessing spherical symmetry.  Any analytical attempts to study realistic
rotating bodies relied on perturbation theory and the assumption of low
angular momentum. It took astrophysicists nearly fifty years after
Einstein first wrote down the final form of his theory to find
an asymptotically flat vacuum solution which has nontrivial
angular momentum (Kerr 1963), and another thirty-three years to
construct appropriate interior solutions (see, e.g., Lynden-Bell
\& Pichon 1996).  Some progress has
been made in the study of relativistically
rotating, axisymmetric objects with finite physical extension,
but mostly in the limit when the disk is
cold (Bardeen \& Wagoner 1971, Lynden-Bell \& Pineault 1978b),
or when the material in it is taken in the form of
two equal, collisionless, counter-rotating sheets
(Lynden-Bell \& Pineault 1978a, Lemos 1989).  The
first case results in the mathematical simplification
that the number of unknown metric functions reduces
to three; the second, in the elimination of the dragging
of inertial frames.

However, the Newtonian analogs to cold
disks are fraught with fierce dynamical instabilities
(see, e.g., Binney \& Tremaine 1987).
Such models cannot represent good approximations for realistic
astrophysical systems (e.g., spiral galaxies, see Bertin \& Lin 1996;
or protoplanetary disks, see Adams \& Lin 1993).  Counter-rotating
disks can avoid instability if they are sufficiently hot,
but since such configurations have to arise as
stellar-dynamical (collisionless) rather than gas-dynamical
(collisional) systems, relativistic analogs may have
difficulty reaching the requisite degree of physical
compactness.

In the Newtonian studies of self-gravitatng disks,
devices that have proven to have great mathematical utility
are the assumptions of complete flattening and self-similarity
(e.g., Mestel 1963, Zang 1976, Toomre 1977, Shu et al. 2000).
Razor-thin disks whose surface densities are power laws
in radius $r$ but which need not possess axial symmetry
have solutions that can be found by
analytical or semi-analytical means (i.e., involving
nothing worse than the numerical integration of ordinary
differential equations [ODEs]; see, e.g., Syer \& Tremaine 1996,
Galli et al. 2001).  The gravitational collapse of such Newtonian
models have elegant self-similar properties in spacetime
(see, e.g., Li \& Shu's [1996] study of the collapse of the
axisymmetric singular isothermal disk).  The relativistic
analogs of such gravitational collapses, axisymmetric and non-axisymmetric,
could lend valuable insight into issues of great contemporary
interest in general relativity, such as the efficiency of
gravitational radiation, or the possible formation of naked singularities.

As preparatory work toward such applications,
we wish to extend Lynden-Bell and Pineault's (1978b)
work on cold, axisymmetric, relativistically rotating disks
to obtain semi-analytical
solutions of a family of relativistic disks parameterized by constant
isothermal sound and rotation speeds, $a$ and $V$.  Self-similarity
is then dimensionally still possible in the relativistic
regime, because relativity introduces only another
constant, the speed of light $c$, with the same units of velocity.
Since no power of $G$, the universal gravitational constant,
combined with $c$ (or $a$ or $V$) can yield a quantity
with the dimensions of length (or time), it becomes natural
and feasible to look for unbounded disk solutions where
the surface density varies as a power of some appropriately
chosen radial coordinate $r$.  The mathematical consequences
of this basic idea are developed in the paper as follows.
Sections \ref{Basic Equation} and
\ref{Final Equation} derive the equations of axisymmetric
stationary spacetime generated by a rotating disk.  Section
\ref{Numerical implementation} describes our numerical strategy
for the solution of the resulting ODEs.  Section \ref{Results}
gives the results of the numerical integrations, and it also
explores some properties of massive test particles placed
in circular orbits in the disk plane.
Section \ref{Photon Orbits} considers the orbits of massless test particles
(photons or neutrinos) in the general spacetimes
of our models.  Finally, in Section \ref{Conclusion}
we offer our conclusions and speculations.

\section{Dimensionless Basic Equations} \label{Basic Equation}

\subsection{Elementary Dimensional Considerations}

We characterize isothermal disks with flat rotation curves by
two dimensionless parameters:
the linear rotation velocity as a fraction of the
speed of light, $v\equiv V/c$, and the
square of the isothermal sound speed
as a fraction of $c^2$,
$\gamma \equiv (a/c)^2$.
We nondimensionalize by adopting the unit
of mass per unit length as
$c^2/G$ = $1.35\times 10^{28}$ g cm$^{-1}$ = 0.677 $M_\odot$ km$^{-1}$.
Notice that with $c$ and $G$ alone, we cannot define a characteristic
mass per unit area (surface density), nor can we define a characteristic
length.  Thus, if $r$ is a coordinate radius, with the physical
units of length, we are naturally interested in disks with
surface densities that are proportional to $c^2/Gr$, i.e.,
with surface densities that are inversely proportional to one
power of $r$.

In general relativity it is possible to accomplish the
equivalent nondimensionalization by working in the geometrical
system of units where $c=G=1$.  It is also possible, of course, to choose
a radial coordinate $r_{\rm LP}$ that does not have the units of length
(see, e.g., Lynden-Bell \& Pineault 1978b and Lemos 1989).
However, to maintain self-similarity in the problem,
$r_{\rm LP}$ can be at best only some power ($1/k$ below)
of $r$.  Using $r_{\rm LP}$ allows us to specify in advance that in
going from pole to pole along a locus $r_{\rm LP}$ = constant
(for a slice at constant $\phi$ and $t$),
the associated polar angle $\theta$ ranges from 0 to $\pi$,
with the disk midplane located by symmetry at $\theta=\pi/2$.
However, this adherence to normal convention comes only at the expense
of making $k$ a nonlinear eigenvalue of the problem.  Using
$r$ and absorbing $k$ into the definition of $\theta$
eliminate the need for a complicated numerical procedure
to find the value of $k$, but it puts the location of the
midplane at a polar angle $\theta = \theta_0 \neq \pi/2$.
The former represents a considerable computational advantage,
whereas the latter serves as a useful reminder that if
we choose a coordinate $r$ such that ${\cal R} \propto r$ represents
the physical radial distance from the origin
to the point $(r,\theta,\phi)$ in the disk midplane
(with $\cal R$ having operational meaning as a proper radius
because the origin is only mildly singular),
then the distance along constant $\cal R$ (again for a slice
at constant $\phi$ and $t$)
from the midplane to the pole may not equal $\pi {\cal R}/2$
because of the distortion of the spatial geometry produced
by the flattened mass distribution.

\subsection{The Metric}

Without losing generality, the metric that is stationary,
axisymmetric and invariant under $\phi \rightarrow -\phi$ and $t
\rightarrow -t$ may be written in geometrical units as
\begin{equation}
    ds^2=-e^{2\nu} dt^2 + B^2 e^{-2\nu} r^2 \sin^2 \theta (d\phi -
\omega dt)^2 + e^{2(\mu-\nu)} (dr^2 + r^2 d\theta^2),
\label{naivespacetime}
\end{equation}
where $\nu$, $B$, $\omega$ and $\mu$ are in general functions of
$r$ and $\theta$.  Since there is no fundamental
unit of length, one might naively
conclude that $\nu$, $B$, $\omega r$ and $\mu$ are functions of
$\theta$ only in a self-similar disk.
This conclusion is premature and false.

In the weak field limit, when the surface density and the rotation
velocity are small, a cold Mestel disk of infinite extent has the associated
gravitational potential,
\begin{equation}
-\nu = \Phi = -v^2 \ln \left[ \frac{r}{D}(1+|\cos \theta|)\right],
\end{equation}
where $D$ is a fiducial length scale that contributes
only an added constant to the potential and thus enters
nowhere else in the problem.  We will discard $D$ in what follows.
In the Newtonian limit, therefore, $e^{\nu}\propto r^n$,
where $n\approx v^2$ when $v^2\ll 1$ (with
$n$ having a different dependence on $v$
and possibly also on $\gamma$ when the rotation and
isothermal sound speeds in the disk are not small
compared to $c$).
This behavior -- the non-predetermined power-law of the
gravitational distortion of time and the spatial geometry by
the disk's self-gravity -- is the
source of the scaling relationships described
by Lynden-Bell \& Pineault (1978b) and Lemos (1989).
According to the nomenclature of Barenblatt (1976),
the situation is an example of self-similarity of
the second kind.

Equation (\ref{naivespacetime}) allows a spacetime
that is self-similar under the transformation
$r\rightarrow \alpha r$, $t\rightarrow \alpha^{1-n} t$.  Then $ds^2
\rightarrow \alpha^{2} ds^2$, and we can write the metric
for $0<\theta<2\theta_0$ as
\begin{equation}
    ds^2 = -r^{2n} e^N dt^2 + r^{2} e^{2P-N} (d\phi - r^{n-1} e^{N-P}
Q dt)^2 + e^{Z-N} (dr^2 + r^2 d\theta^2), \label{metric}
\end{equation}
where $N$, $P$, $Q$ and $Z$ are functions of $\theta$ only.
In the above metric, $n$ is a pure number between 0 and 1
that measures the depth of the disk's gravitational well.
In particular, photons that emerge from the origin
and reach some finite $r$ have frequencies that are
infinitely redshifted from their starting values.
Self-similarity then implies that the same infinite
redshift applies to photons that originate at any finite
$r$ and try to propagate to infinity (see \S 5).
In a certain sense, therefore, the system
constitutes an incipient black hole, one that
will presumably acquire a growing point mass at the
origin, with an accompanying horizon, if the disk
is unstable and undergoes inside-out gravitational collapse
(see Li \& Shu 1996 for a Newtonian analog).

It is instructive to compare our metric to the one used by
Lynden-Bell \& Pineault (1978b) and Lemos (1989):
\begin{equation}
    ds^2 = -r_{\rm LP}^{2m} e^N dt^2 + r_{\rm LP}^{2k} e^{2P-N} (d\phi - r_{\rm LP}
^{m-k} e^{N-P} Q dt)^2 + r_{\rm LP}^{2(k-1)} e^{Z_{\rm LP}-N} (dr_{\rm LP}^2 + r_{\rm LP} ^2
d\chi^2).
\end{equation}
The relationship between the two coordinate conventions
can be made explicit by the following transformation:
\begin{equation}
    r = r_{\rm LP}^k, \qquad \theta = k \chi, \qquad n = \frac{m}{k}, \qquad
Z = Z_{\rm LP} - \ln k.
\end{equation}
While the two metrics are completely equivalent, our choice turns
out to be more advantageous in numerical implementation.  For Lemos and
Lynden-Bell \& Pineault, the disk is located at $\chi = \pi/2$, and $m$
and $k$ are eigenvalues of the problem.  To find their values and
the scaling of rotation velocity with density and pressure
requires a three-dimensional shooting method, a nontrivial
numerical task.  For our metric,
we have only one eigenvalue $n$.  The other degree of freedom is
embedded in the location of the disk, which we can find
by satisfying certain jump conditions when we cross
from the top to bottom surface.  Later on, we will
reparameterize the solution space to avoid even having to find
$n$ as an eigenvalue.

Define the orthonormal tetrads for the locally non-rotating
observer (Lemos 1989):
\begin{equation}
    \tet{0}{\mu} = (r^{-n} e^{- N/2}, r^{-1}Q e^{
N/2-P}, 0, 0),
\end{equation}
\begin{equation}
    \tet{1}{\mu} = (0, r^{-1}e^{N/2-P}, 0, 0),
\end{equation}
\begin{equation}
    \tet{2}{\mu} = (0, 0, e^{(N-Z)/2}, 0),
\end{equation}
\begin{equation}
    \tet{3}{\mu} = (0, 0, 0, r^{-1}e^{(N-Z)/2}).
\end{equation}
The indices in parentheses label the basis vectors in
$(t,\phi,r,\theta)$, and are raised
and lowered with the flat Minkowski metric $\eta = {\rm diag} (-1, 1,
1,1)$. The nontrivial Ricci components are
\begin{equation}
    2\R{0}{0}r^2 e^{Z-N}=N_{\theta \theta} +N_\theta P_\theta
    +2n(1+n) - Q^2\left\{\left[ \left( \ln
Q\right)_\theta-P_\theta+ N_\theta\right]^2 + (1-n)^2\right\},
\end{equation}
\begin{equation}
    2\R{0}{1}r^2 e^{Z-N} = Q_{\theta \theta} + Q_\theta P_\theta
     - Q\left[ P_{\theta \theta}-N_{\theta\theta}
    +(P_\theta-N_\theta)^2 + P_\theta(P_\theta-N_\theta) + 2(1-n)\right],
\end{equation}
\begin{equation}
    (\R{0}{0} - \R{1}{1}) r^2 e^{Z-N} = P_{\theta \theta} +
    P_\theta^2 +(n+1)^2,
\end{equation}
\begin{equation}
    2\R{2}{2} r^2 e^{Z-N} = N_{\theta \theta} -Z_{\theta \theta}
    + P_\theta(N_\theta-Z_\theta)+2n(1-n)+Q^2 (1-n)^2,
\end{equation}
\begin{equation}
    2\R{2}{3} r^2 e^{Z-N} = (n+1) Z_\theta -2n N_\theta +
    Q^2 (1-n) \left[ P_\theta- \left(\ln Q\right)_\theta - N_\theta
\right],
\end{equation}
\begin{equation}
    2\left[ \R{3}{3}+\R{0}{0} -\R{1}{1}-\R{2}{2}\right] r^2 e^{Z-N} =
    2P_\theta Z_\theta - N_\theta^2 + 4n^2 +Q^2
    \left\{\left[N_\theta + \left(\ln Q\right)_\theta
- P_\theta\right]^2-(n-1)^2\right\}.
\end{equation}

\subsection{Matter}
For the stationary thin disk, there is no
radial or vertical motion. Hence we may write the four velocity as
\begin{equation}
    u^\mu = r^{-n} e^{-N/2} (1-v^2)^{-1/2} (1, \Omega, 0, 0),
\end{equation}
where
\begin{equation}
    \Omega = \frac{d\phi}{dt} = \frac{u^\phi}{u^t}
\end{equation}
is the coordinate angular velocity and
\begin{equation}
    v = r^{1-n} e^{P-N} \Omega - Q
\end{equation}
is the linear velocity of the fluid in the locally nonrotating
frame. The physical significance of this quantity become clear
when we project the four-velocity onto
the locally nonrotating frame defined by
the tetrad:
\begin{equation}
    u^{(a)} = (\frac{1}{\sqrt{1-v^2}}, \frac{v}{\sqrt{1-v^2}}, 0, 0),
\end{equation}
i.e., the physics is that of special relativity in this frame,
with $v$ equal to a constant. In
order to keep the disk infinitesimally thin, we
will adopt the matter content of a fluid with vanishing vertical
pressure.  In the rest frame of the
fluid, the stress energy tensor is given by
\begin{equation}
    \T{a}{b} =\left( \matrix{
    \varepsilon&0&0&0\cr
    0&p_\phi&0&0\cr
    0&0&p_r&0\cr
    0&0&0&0\cr}\right)\delta(\theta-\theta_0) .
\label{stressenergytens}
\end{equation}
Boosting into the tetrad frame, the nonvanishing components are
(suppressing the argument of the delta function)
\begin{equation}
    \T{0}{0} = \frac{\varepsilon + p_\phi v^2}{1-v^2}\delta,
\end{equation}
\begin{equation}
    \T{0}{1} =
-\frac{(\varepsilon+p_\phi)v}{1-v^2}\delta,
\end{equation}
\begin{equation}
    \T{1}{1} = \frac{p_\phi+ \varepsilon v^2}{1-v^2}\delta,
\end{equation}
\begin{equation}
 \T{2}{2} =  p_r\delta .
\end{equation}
The Einstein field equations take the usual form
\begin{equation}
    \R{a}{b} = 8\pi \left[ \T{a}{b} - {1\over 2} \eta_{(a)(b)} T\right]
\end{equation}
where $T =T^{(a)}_{\phantom{(a)}(a)} = (-\varepsilon + p_\phi + p_r)\delta$
is the trace of the stress-energy tensor.

The equation of motion on the disk is given by the vanishing
divergence of the stress-energy tensor,
$T_{\phantom{\mu\nu};\nu}^{\mu \nu} = 0$.  For this particular
configuration, only the $\mu=r$ component is nontrivial.  The
$\mu= \theta$ component is proportional to the derivatives of the
metric coefficients, which are discontinuous at the equatorial
plane.  By symmetry, this equation is identically satisfied once
integrated through the plane.  After some algebra, the $r$
component reads
\begin{equation}
    n\frac{\varepsilon  + p_\phi v^2}{1-v^2} - \frac{p_\phi
+ \varepsilon v^2}{1-v^2} - Q_0v(1-n)\frac{\varepsilon + p_\phi}
{1-v^2}+ (n+2) p_r + r{dp_r\over dr} =0
\end{equation}
where $Q_0$ is the value of $Q$ on the disk.  With $Q_0$, $n$, and
$v$ equal to constants, this equation
implies that $\varepsilon$, $p_\phi$, and $p_r$ all have the same
radial dependence, which on dimensional grounds must be
$r^{-2}$, as can also be seen from their
coupling to geometry through the Einstein
field equations.  Therefore the above equation becomes an algebraic one:
\begin{equation}
    Q_0 {v(1-n)(\varepsilon + p_\phi)} + p_\phi + \varepsilon v^2
    - n(\varepsilon + p_\phi v^2) - n p_r (1-v^2)=0.
\label{eom}
\end{equation}

\section{Final Form of the Equations}
\label{Final Equation}
\def\et{\tilde \varepsilon}
\def\pft{\tilde p_\phi}
\def\prt{\tilde p_r}
Define the reduced energy and stresses:
\begin{equation}
    \et = 8\pi \frac{\varepsilon}{1+n} r^2 e^{Z_0-N_0}, \qquad \pft = 8 \pi
\frac{p_\phi}{1+n} r^2 e^{Z_0-N_0}, \qquad \prt = 8\pi \frac{p_r}{1+n} r^2
e^{Z_0-N_0},
\label{tilde}
\end{equation}
where a subscript $0$ denotes the value on the disk at $\theta =
\theta_0$. Define a further rescaling,
\begin{equation}
\Theta \equiv (1+n) \theta,
\end{equation}
and let a prime denote differentiation with respect to
$\Theta$.  Coupled to matter, the equation for $P$ takes the form
\begin{equation}
    P''+ {P'}^2 + 1 =
    \prt \delta (\Theta - \Theta_0)
\label{eqP1}
\end{equation}
Away from the disk, we can solve this equation subject to the
boundary condition $e^{P(0)} = 0$ (so that a circle around
the axis will have vanishingly small circumference as $\Theta
\rightarrow 0$):
\begin{equation}
    P = \ln \left[ \sin \Theta\right] + C, \qquad
    P' = \cot \Theta.
\label{Psoln}
\end{equation}
The constant $C$ remains arbitrary, which enables us to set the
boundary condition for other metric functions later.  The
solution (\ref{Psoln}) is only valid in the range
$0<\Theta<\Theta_0$ where $P$ is
differentiable.  For $\Theta>\Theta_0$, we can obtain the
solution simply through symmetry considerations
(i.e., the metric functions are
even about the disk, while the derivatives are odd). Integrating
equation (\ref{eqP1}) across $\Theta = \Theta_0$ and
combining the result with the second relation of
equation (\ref{Psoln}), we get
\begin{equation}
    -2\cot \Theta_0 = \prt. \label{pr}
\label{plane}
\end{equation}

Let us confine our attention to $0 \le \Theta \le \Theta_0$.
The rest of the field equations become
\begin{equation}
    N'' + N' P' + \frac{2n}{1+n} - Q^2
\left\{\left[\left(\ln Q\right)' - P' + N'\right]^2+
\left(\frac{1-n}{1+n}\right)^2\right\}=
\left[\prt+(\et +
\pft)\frac{1+v^2}{1-v^2}\right]\Delta ,
\end{equation}
\begin{equation}
    Q'' + Q' P' - Q\left[\frac{2(1-n)}{(1+n)^2} -1
    -N''+ (N' - P')^2 - \cot
    \Theta N'\right] = \left[Q\prt -2 (\et+\pft) \frac{v}{1-v^2}\right]
\Delta,
\end{equation}
\begin{equation}
    Z'-\frac{2n}{1+n} N'+ Q^2\frac{1-n}{1+n}
\left[ P' -\left(\ln Q\right)'-N'\right] = 0,
\end{equation}
\begin{equation}
    Q^2\left\{\left[N' + \left(\ln Q\right)' - P'\right]^2-\left(
\frac{1-n}{1+n}\right)^2\right\} +
    2 P' Z' - {N'}^2 + \frac{4n^2}{(1+n)^2} = 0.
\end{equation}
where $\Delta = \delta(\Theta - \Theta_0)= \delta(\theta -
\theta_0)/(1+n)$.  Here we have dropped the equation that involves
$Z''$.  It can be recovered by the equation of motion (\ref{eom})
owing to the contracted Bianchi identity. These equations can be
simplified to give the dynamic equations,
\begin{equation}
    N'' + N' P' + \frac{2n}{1+n} -
\left( Q' - Q P' + Q N'\right)^2- Q^2 \left( \frac{1-n}{1+n}\right) ^2=
\left[\prt+(\et + \pft)\frac{1+v^2}{1-v^2}\right]\Delta ,
\label{dyneqN}
\end{equation}
\begin{equation}
    Q'' + Q' P' - Q\left[
        (1-Q^2) \left( \frac{1-n}{1+n}\right)^2 + (N'-P')^2
    -\left( Q' - Q P' + Q N'\right)^2
    \right]
    = -(\et+\pft) \frac{2v+Q+Qv^2}{1-v^2}\Delta .
%    &Z''+ Z' \cot \Theta - 2Q^2\prn{\frac{k-n}{k+n}}^2
%+ \frac{4n^2}{(k+n)^2} - \prn{Q' - Q \cot \Theta + Q
%N'}^2 = 2\brk{\frac{\et v^2 + \pft}{1-v^2}} \Delta
\label{dyneqQ}
\end{equation}
and the constraint equation,
\begin{equation}
    \left[Q P' - Q' - Q N'\right]^2 - Q^2
    \left(\frac{1-n}{1+n}\right)^2 + \frac{4n}{1+n} P' N'
    - 2 Q
    P' \frac{1-n}{1+n} \left[Q P' - Q' - Q N'\right] -{N'}^2 +
    \frac{4n^2}{(1+n)^2} =0 .
%    &Z'-\frac{2n}{n+k} N'+ Q\frac{k-n}{k+n} \brk{Q \cot \Theta -
%Q'- QN'} = 0\\
%    &\prn{Q N' + Q' - Q\cot \Theta}^2- Q^2 \prn{\frac{n-k}{n+k}}^2 + 2
%    \cot \Theta Z' - {N'}^2 + \frac{4n^2}{(n+k)^2} = 0
\label{coneq}
\end{equation}
Now, $Z$ has been completely decoupled from $Q$ and $N$.  We may
use equation (\ref{coneq}) to decouple $Q$ from $N$ as well, but
this offers little advantage for numerical purposes.

\section{Numerical Implementation}
\label{Numerical implementation}
\subsection{Boundary conditions}
We have already discussed the boundary condition for $P$.  Unless
$r=0$, we expect space to be regular on the axis, which means the
Ricci tensor remains finite there (and thus Riemann normal coordinates
exist there).  On the pole, $P' = \cot \Theta$
diverges as $\Theta^{-1}$. In order to have a regular solution for
equation (\ref{coneq}) there, we require
\begin{equation}
    N' = 0, \qquad Q = 0 \qquad {\rm at} \qquad \Theta = 0.
\end{equation}
The first requirement prevents the $\Theta$ geometry from having
a cusp at the pole; the second discounts frame dragging on the
rotation axis.  One might naively expect $Q'=0$ on the pole as well from
the term $Q'P'$ in  equation (\ref{dyneqQ});  however, this
singularity is cancelled by $Q {P'}^2$.  In fact, we can use $Q'$
on the axis to parameterize the solution space . With proper
rescaling of $t$, $r$ and $e^P$, we can set $N = 0$ on
the pole (which means $r^n \, dt$ is the interval
of proper time of an observer at $\Theta=0$).

On the other hand, the delta functions on the right hand sides
of the governing ODEs (\ref{dyneqN}) and (\ref{dyneqQ})
signal a discontinuity in the derivatives of metric
coefficients when we cross the plane of the disk.
Similar to the method we used to find $P$, we
integrate across the disk, and assume that the geometry is
symmetric about the disk plane.  Equations (\ref{dyneqN})
and (\ref{dyneqQ}) then yield the following boundary
conditions at $\Theta=\Theta_0$:
\begin{equation}
    N'_0 = - \frac{1}{2} \left[\prt+ (\et + \pft)
    \frac{1+v^2}{1-v^2}\right], \qquad
    Q'_0 = \frac{1}{2} (\et+\pft) \frac{Q_0+Q_0v^2+2v}{1-v^2}.
%    &Z'_0 = \frac{1}{n+k}\frac{\et v^2+\pft}{1-v^2}\\
\label{jump}
\end{equation}

\subsection{Scalar Two-Dimensional Pressure and Method of Solution}
For simplicity, we adopt a planar isotropic equation of state,
\begin{equation}
    p_r = p_\phi = \gamma \varepsilon,
\end{equation}
where $\gamma$ is the isothermal sound speed squared, as usual.  Then,
on the disk, equations (\ref{eom}), (\ref{plane}), and (\ref{jump}) imply
\begin{equation}
    Q_0 v(1-n)(1+\gamma) + \gamma + v^2 - n (1+\gamma) =
0,
\label{eom1}
\end{equation}
\begin{equation}
    N_0' = -\frac{\et}{2} \left[\frac{2\gamma + 1 +v^2}{1-v^2}
\right],
\label{NP}
\end{equation}
\begin{equation}
    Q_0' = \frac{\et}{2}\left(1+\gamma\right) \left[\frac{Q_0(1+v^2) +
2v}{1-v^2}\right].
\label{QP}
\end{equation}
\begin{equation}
   \et \gamma = -2\cot \Theta_0.
\label{epsgam}
\end{equation}

We may give equation (\ref{eom1}) the following quasi-Newtonian
interpretation.  Because of nonzero pressure, the effective
gravitational mass-energy is enhanced by the factor $(1+\gamma)$
for both the effects of gravitation $n$ and the dragging of
inertial frames $Q_0v(1-n)$; the net effect of these two terms is
balanced per unit inertial mass-energy by the ``centrifugal term''
$v^2$ (with the ``radius'' scaled out in this self-similar
problem) and the pressure term $\gamma$.  Similarly,  equation
(\ref{NP}) is the analog of the Newtonian relationship that the
vertical gravitational field ($\propto N_0'$) just above the
surface of the disk is equal to $-2\pi G$ times the local surface
mass-density ($\propto \et$); the extra factor represents various
relativistic corrections (see the first relation of eq.
[\ref{jump}]).  The jump condition (\ref{QP}) on the derivative of
the frame-dragging term $Q_0'$ and the geometrical distortion
(\ref{epsgam}) of the angular location of the midplane of the disk
have, of course, no Newtonian analogs.

For a cold, slowly rotating, disk, where
$\gamma \ll v^2 \ll 1$, the frame-dragging term $\propto Q_0v$
is negligible, and equation (\ref{eom1})
recovers the Newtonian approximation for a centrifugally
supported Mestel disk: $n\approx v^2$.  On the other hand,
from equation (\ref{Nprimeprime}) below, $N'$ in this limit
has a solution which satisfies $N'= 0$ at $\Theta = 0$ given by
\begin{equation}
N' \approx -{2n\over 1+n}\left( {1-\cos\Theta\over \sin\Theta}\right),
\end{equation}
with equation (\ref{epsgam}) yielding the location of the
disk midplane at $\Theta_0 \approx \pi/2$.  Thus, $N_0'
\approx -2n \approx -2v^2$, and equation (\ref{NP}) now leads
to the solution, $\et \approx 4v^2$, where $\et$
is itself obtained from the first relation of equation (\ref{tilde})
as $\et \approx 8\pi {\cal R}^2 \varepsilon$ (see eq. [\ref{calR}]
below).  Thus, we have the
identification $\varepsilon \approx v^2/2\pi {\cal R}^2$, which
corresponds to a Newtonian surface-mass density $\Sigma =
c^2 {\cal R}\varepsilon/G$ (radius $\cal R$ now having the dimensions of
length) related to the disk rotational velocity
$V=cv$ given by the famous Mestel formula:
\begin{equation}
\Sigma = {V^2\over 2\pi G {\cal R}}.
\end{equation}

For the fully relativistic situation, the disks are characterized by
values of $v$ and $\gamma$ that are not very small compared
to unity.  In such an situation, one approach could be to
specify these two parameters and solve the problem numerically with $n$ and
$\Theta_0$ as eigenvalues. In practice, such an approach is very costly.
We would have to adopt a shooting method in three dimensions for $n$,
$\Theta_0$ and $\eta = Q'(0)$.  For nonlinear ODEs,
the number of operations increases exponentially with
the number of eigenvalues.  On the other hand, the
values of $v$ and $\gamma$ do not come into play until we get to
the disk because the right-hand sides of equations
(\ref{dyneqN}), (\ref{dyneqQ}), and (\ref{coneq})
vanish when $\Theta \neq \Theta_0$. Therefore,
it is more efficient to treat $n$ and $\eta$ as our nominal
solution-space parameters, and solve for $\Theta_0$, $v$,
$\gamma$, and $\et$ from equations (\ref{eom1}), (\ref{NP}),
(\ref{QP}), and (\ref{epsgam}) once we have integrated the ODEs
(\ref{dyneqN}), (\ref{dyneqQ}), and (\ref{coneq})
for the properties of spacetime off the disk plane.

The quantities $\Theta_0$, $v$ $\gamma$, and $\et$
enter in equations (\ref{eom1})--(\ref{epsgam}) with a pattern that
allows us to proceed as follows.
The dynamic equations (\ref{dyneqN}), (\ref{dyneqQ}), (\ref{coneq}) are
cast for $\Theta \neq \Theta_0$
as a set of two second-order ODEs in $N$ and $Q$:
\begin{equation}
    N'' = Q^2\left(\frac{1-n}{1+n}\right)^2 + \left(Q' - Q \cot \Theta +
QN'\right)^2 - \frac{2n}{1+n} - N' \cot \Theta,
\label{Nprimeprime}
\end{equation}
\begin{equation}
    Q'' = Q\left[(1-Q^2)\left(\frac{1-n}{1+n}\right)^2 +
(N'-\cot \Theta)^2 -
(Q'- Q\cot \Theta + QN')^2\right] - Q' \cot \Theta .
\end{equation}
For given $n$ and $\eta\equiv Q'(0)$, we begin with the boundary
values, $Q=0$, $Q'=\eta$, $N=0$, $N'=0$, at the pole $\Theta =0$
and integrate toward the disk at $\Theta=\Theta_0$ (whose value
is unknown at this point).
At each potential choice $\Theta$ for $\Theta_0$,
we solve equation (\ref{eom1}) for $1+\gamma$:
\begin{equation}
    1+\gamma = \frac{1-v^2}{(Q_0 v + 1)(1-n)} \label{g+1},
\label{onepgam}
\end{equation}
where $v$ is obtained by the following procedure.
We first divide equation (\ref{QP}) by equation (\ref{NP}):
\begin{equation}
    A \equiv -\frac{Q_0'}{N_0'} = \frac{(1+\gamma)\left[Q_0(1+v^2) + 2v\right]}
{2 (1+\gamma) -(1-v^2)},
\end{equation}
with the value of $A$ known from the off-plane integration
to the candidate $\Theta$ for $\Theta_0$.
With the elimination of $1+\gamma$ through equation
(\ref{onepgam}), the last equation implies
\begin{equation}
Q_0 v^2 + \left[2+AQ_0 (1-n)\right] v + Q_0 - A(1+n)=0.
\end{equation}
This quadratic equation for $v$ yields a solution,
\begin{equation}
    v = -\frac{1}{Q_0} - \frac{A}{2}(1-n) +
\sqrt{\left[ \frac{1}{Q_0} + \frac{A}{2} (1-n)\right]^2 - 1 +
\frac{A}{Q_0} (1+n)},
\end{equation}
where we have chosen the sign so that $v$ is well behaved,
going to $(1+n) A/2 \rightarrow 0$, when $Q_0 \rightarrow 0$.

Once $v$ and $\gamma$ are known, we can
calculate the rescaled energy density
\begin{equation}
    \et = - \frac{2(1-v^2) N_0'}{1+2\gamma + v^2}.
\end{equation}
We have found the disk when $\Theta$ has a value
$\Theta_0$ such that equation (\ref{epsgam}) holds:
\begin{equation}
    \cot \Theta_0 = -{1\over 2}\gamma \et.
\end{equation}
Since the reduced energy density $\et$ is positive,
we obtain a disk location $\pi/2 <\Theta_0<\pi$.
With the other obvious limits,
\begin{equation}
    0<v<1, \qquad 0<\gamma<1,
\end{equation}
our parameter space is confined to
\begin{equation}
    0<n<1, \qquad \eta > 0.
\end{equation}

\section{Results}
\label{Results}

\begin{figure}[ht]
\begin{center}
\includegraphics[width = 4in, angle = 90]{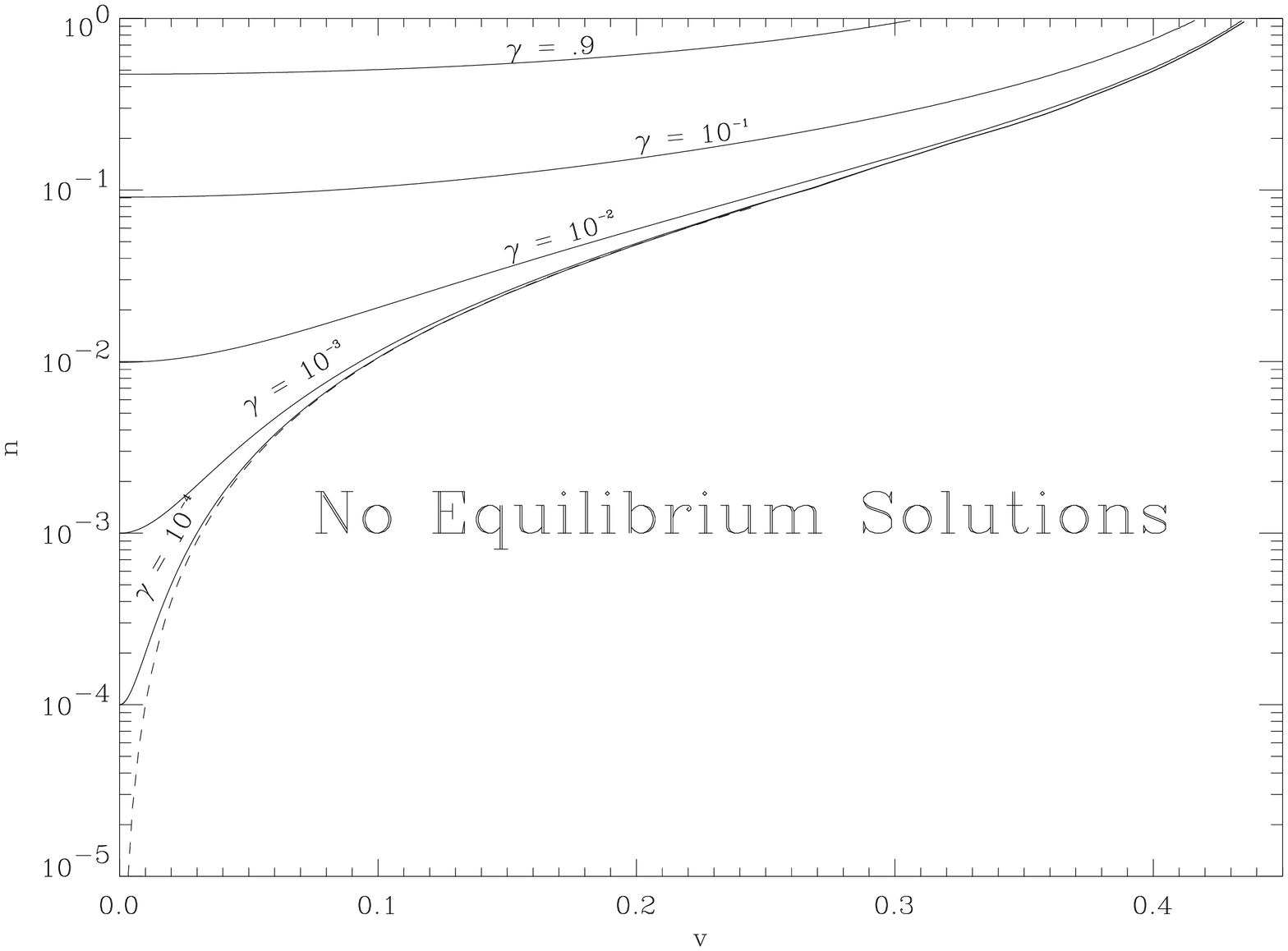}
\caption{The gravity index $n$ vs rotational velocity $v$ for
different values of the pressure parameter $\gamma$.  The dashed line
is the empirical approximation to $n$ for $\gamma = 0$, found by
Lynden-Bell and Pineault.}
\label{constg}
\end{center}
\end{figure}

Contours of constant $\gamma$ in $n-v$ space are plotted in Figure
\ref{constg}.  Roughly speaking, $n$ is a measure of the strength
of the gravitational field.  It ranges from $n=0$ for flat space
to $n=1$ for maximum rotation.  One may also argue that since our
similarity transformation is $r\rightarrow \alpha r$, $t
\rightarrow \alpha^{(1-n)}t$, $n$ cannot exceed unity or the
passage of time would proceed more quickly deeper in the
gravitational potential well, contrary to common experience in
general relativity. Hence, $n$ lying within the interval (0,1) is
an anticipatable result from the self-similarity of the basic
problem. From Figure \ref{constg} we also see that equilibria
require lower $v$ for given $n$ if $\gamma$ is larger; this result
conforms with the Newtonian intuition that less rotational
velocity is needed in a disk to offset the self-gravity if there
is a greater degree of pressure support.

For each $\gamma$, there is a maximum rotational velocity $v_c$
above which there is no equilibrium.
Table 1 gives the numerical value of $v_c$
as a function of the parameter $\gamma$.  Our computed value $v_c=0.436$ when $\gamma = 0.004$ is consistent,
if we perform a simple extrapolation,
with Lynden-Bell and Pineault's (1978b) estimate, $v_c=0.438$
when $\gamma = 0$, and the latter numbers are what are entered
as the first entry of Table 1.

\begin{figure}[ht]
\begin{center}
\includegraphics[width = 4in, angle = 90]{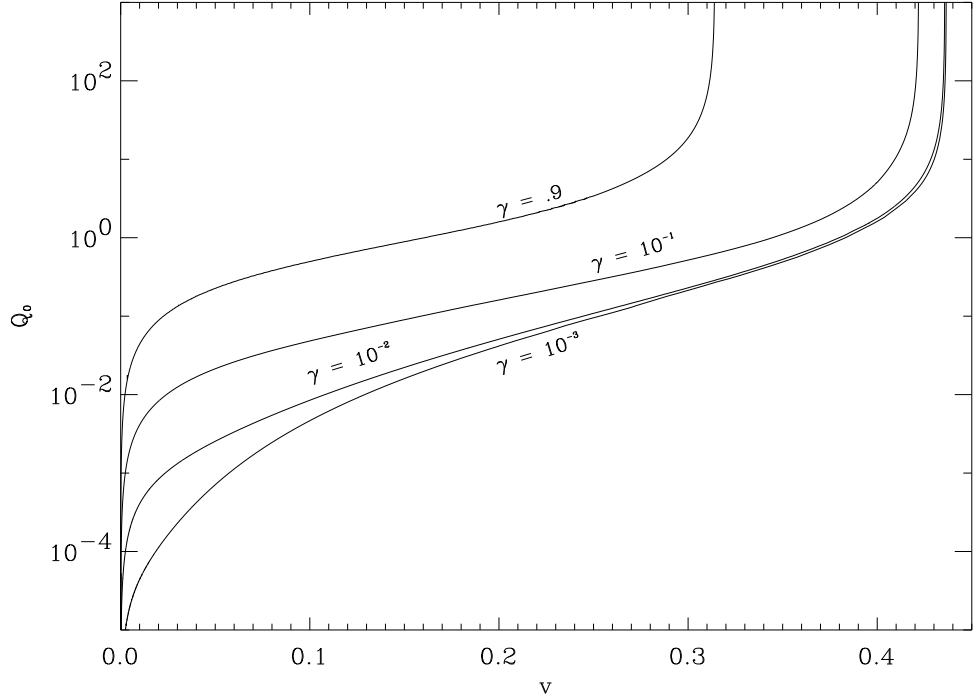}
\caption{The frame dragging parameter $Q_0$ vs rotational
velocity $v$ for different values of the pressure parameter $\gamma$.}
\label{qv}
\end{center}
\end{figure}

To understand the last result physically,
we examine in Figure \ref{qv}, for various choices of
$\gamma$, the behavior of the term
$Q_0$ that governs the dragging of inertial frames in the disk plane.
Nonzero values of $Q_0$
represent how hard one must accelerate to remain at constant
$\phi$.  In our metric, $g_{tt} \propto (1-Q^2)$.  Thus, whenever
$Q>1$, an ergoregion develops and the time-like Killing vector
$\partial/\partial t$ becomes space-like.
Since $Q$ only has dependence on
$\theta$, the ergoregion is best described by the exterior of a cone
whose opening angle $\theta_{\rm{ergo}}$ is defined by
$Q(\theta_{\rm{ergo}}) = 1$.  Naturally, the ergoregion first
appears on the disk when $Q_0=1$ and $\theta_{\rm{ergo}} = \theta_0$.
If we assume $Q$ is continuous and monotonic, then
$\theta_{\rm{ergo}}$ decreases (for the ergoregion above the disk) as
$Q_0$ increases, until finally $\theta_{\rm{ergo}} \rightarrow 0$ as
$Q_0 \rightarrow \infty$.  In this limit, when the ergoregion
occupies the entire space above (and below) the disk,
equation (\ref{eom1}) may be balanced
only when $n\rightarrow 1$, which recovers the upper bound on $n$.
The rotation velocity at which $Q_0$ diverges is then
determined by a limit process on the product $Q_0(1-n)$.

\subsection{Surface Density of Models}

Table 2 gives the dimensionless coefficient
\begin{equation}
\hat {\cal E} \equiv {(1+n)\et\over 8\pi},
\label{Ehat}
\end{equation}
(multiplied by 100 to avoid writing too many zeroes)
corresponding to a given pair of values $v/v_c$ and $\gamma$
that characterizes an equilibrium model.  Notice that
for small $\gamma \ll (v/v_c)^2\ll 1$, we recover the
Mestel solution, $\hat {\cal E} \approx v^2/2\pi$.
In principle, if one takes the attitude that gas pressures
must be three-dimensionally isotropic rather than two-dimensionally
as idealized in this paper, then disks cannot remain vertically
thin unless $\gamma \ll v^2$.  In these restricted circumstances,
the entries in Table 2 that violate this constraint are
not physically self-consistent.  Under a broader interpretation of
what might be acceptable in the physical world,
relativistic SIDs might be constructed from non-interacting
dark matter particles,
in which case allowable stress tensors include diagonal
forms that are non-isotropic in the sense of equation
(\ref{stressenergytens}).  In the Newtonian
regime, it is also known that strongly magnetized,
self-gravitating disks, can be vertically thin even
at zero rotation speeds (Shu \& Li 1997 and references therein).
Since there may be useful relativistic analogs of such
systems, an open attitude would retain all the entries
in Table 2 for the sake of mathematical completeness.

In terms of $\hat {\cal E}$, the dimensional surface density
of mass-energy $\cal E$, in units of mass per unit area,
as would be measured by a corotating observer, is obtained from
\begin{equation}
{\cal E} = {c^2\over G} \int_0^{2\theta_0} \varepsilon \delta
(\theta-\theta_0) e^{(Z-N)/2}\, rd\theta =
\left( {c^2\over G}\right){\hat {\cal E}\over {\cal R}},
\label{surfdensity}
\end{equation}
where $\cal R$ is the distance measured from the origin
to the point in question along a radial slice at
constant $\phi$ and $t$ in the equatorial plane
$\theta = \theta_0$ of the disk,
\begin{equation}
{\cal R} = e^{(Z_0-N_0)/2}r.
\label{calR}
\end{equation}
In equation (\ref{surfdensity}), $\cal E$ is given
in $M_\odot$ km$^{-2}$ when $c^2/G$ is taken to be
0.677 $M_\odot$ km$^{-1}$ and $\cal R$ is measured in km.
It should be noted (a) that a proper radius $\cal R$
can be defined in the present
circumstance because the origin neither contains a mass point
nor is shielded from an observer at $\cal R$ by a horizon,
and (b) that the circumference $\cal C$ of a circle
in the plane of the disk according to a nonrotating observer
at radius $\cal R$ is {\it not} given by $2\pi {\cal R}$,
but by
\begin{equation}
{\cal C} = e^{P_0-N_0/2}\, 2\pi r
= e^{P_0-Z_0/2}\, 2\pi {\cal R}.
\end{equation}

\subsection{Circular Orbits: Existence and Stability}

We anticipate that many of the more slowly rotating members of the
singular isothermal disks (SIDs) studied here are unstable to
inside-out gravitational collapse in a similar way as their
Newtonian counterparts (Li \& Shu 1997, Shu et al. 2001). We leave
for a future endeavor the study of the dynamical,
self-gravitating, stability of relativistic SIDs and the
consequent formation of black holes at their centers if they
undergo gravitational collapse.  Here we ask the simpler question:
Are relativistic SIDs {\it kinematically} stable in the sense of
having stable circular orbits for (noninteracting dark-matter)
test particles of nonzero rest mass?  The question is nontrivial,
because circular orbits of arbitrary sizes around point masses in
the Newtonian case are all stable, yet circular orbits lose their
stability if they approach too closely the event horizons of
relativistic point masses (Schwarzschild or Kerr black holes). Is
there a similar loss of orbit stability, when we go from Newtonian
SIDs to relativistic SIDs?

For a test particle of mass $m$ in the equatorial plane of the
disk, the symmetries of the geometry gives the conserved
quantities:
\begin{equation}
    u_t = -\Et = -E/m, \qquad u_\phi = \lt=l/m.
\end{equation}
Since $u^\theta = 0$ for an orbit confined to the disk plane, the
geodesic equation takes the simple form
\begin{equation}
   \left({dr\over d\tau}\right)^2  = g^{rr}[-1 -\Et^2g^{tt} +2\Et\lt g^{t\phi} -
\lt^2
    g^{\phi\phi}] \equiv - 2V(r),
\label{orbitODE}
\end{equation}
where $\tau$ is the proper time of the particle and $V(r)$ is the
effective potential of the problem:
\begin{equation}
  V(r) \equiv {1\over 2} e^{-Z_0}[e^{N_0} -r^{-2}(\Et r^{1-n}-Q_0\lt e^{N_0-P_0})^2
    +\lt^2 r^{-2} e^{2N_0-2P_0}] ,
\end{equation}
and $N_0$, $P_0$, $Q_0$, and $Z_0$ have simple fixed numerical
values when $N$, $P$, $Q$, and $Z$ are evaluated in the disk plane
$\theta=\theta_0$.

To have a circular orbit, we need $dr/d\tau = d^2r/d\tau^2 = 0$,
which implies $V(r) = V^\prime (r) = 0$. These two conditions
define the values of specific energy and angular momentum, $\Et$
and $\lt$, needed to yield a circular orbit at radius $r$.  A
little algebra shows that the required values of $\lt$ and $\Et$
for a circular orbit with $r=r_0$ are
\begin{equation}
    \lt = \pm \sqrt{F-1}e^{P_0-N_0/2}r_0, \qquad \Et = \left[ \sqrt{F} \pm
    Q_0\sqrt{F-1}\right]e^{N_0/2}r_0^n,
\label{param-disk}
\end{equation}
where we have defined
\begin{equation}
   F \equiv
    \frac{2/(1-n)-Q_0^2\mp Q_0\sqrt{Q_0^2+4n/(1-n)^2}}{2(1-Q_0^2)}.
\label{Fdef}
\end{equation}

In equations (\ref{param-disk}) and (\ref{Fdef}), the upper
sign choice corresponds to prograde orbits; the lower, to retrograde
ones.  For the upper sign choice,
the numerator in equation (\ref{Fdef}) goes through zero when the
denominator does; i.e., for any allowable $n$, $F$ stays positive
when the disk becomes an ergoregion as $Q_0$ crosses unity.
>From our earlier discussions, we recognize the factor $r_0^n$
as the correct scaling to account for the contribution to $\Et$ of
the gravitational ``potential energy'' per unit mass.  We
shall prove below that $F>1$ and that the term $\sqrt{F}$ represents the
contribution of the rest and kinetic energies.  Notice now
that frame dragging adds a
positive contribution to the specific energy $\Et$ of prograde
circular orbits, whereas it adds a negative contribution
for retrograde circular orbits.  Moreover, with the
lower sign choice in equation (\ref{Fdef}), $F$ diverges to
$+\infty$ when $Q_0 \rightarrow 1-$; i.e., the two terms in
$\Et$ cancel to lowest order for large $F$
when the disk plane first becomes an ergoregion.  Massive
test particles in retrograde circular motion have zero
total energy in this limit.

The physical interpretation of these results follows from
examining the three-velocity of circular orbits in the tetrad frame:
\begin{equation}
    v_p = \frac{u^{(\phi)}}{u^{(t)}} =\pm \sqrt{\frac{F-1}{F}},
\label{partvel}
\end{equation}
where, again, the upper choice corresponds to prograde motion;
the lower choice, to retrograde motion.
Equation (\ref{partvel}) shows that $F$ is the square of the
Lorentz factor of the particle motion in the tetrad frame:
$F = 1/(1-v_p^2)$.  The first term $\sqrt{F}$ in the expression
for $\Et$ thus represents the usual special-relativistic
contribution to the rest and kinetic energies of a particle.
Moreover, the expression for specific angular momentum
$\lt$ in equation (\ref{param-disk}) is now recognized as
the velocity $v_p$ in the $\phi$ direction,
times the usual Lorentz factor correction,
times, not the radius ${\cal R}_0$, but the
circumference of the orbit ${\cal C}_0$ divided by $2\pi$.
The above interpretation for $F$ explains why
the sequence of retrograde circular orbits terminates
when $Q_0\rightarrow 1$: When the disk develops
an ergoregion, massive particles in retrograde motion
must travel at the speed of light ($F\rightarrow \infty$)
if they are to resist the dragging of inertial frames.

No such difficulty affects particles in prograde circular orbit.
Equations (\ref{param-disk}) and (\ref{partvel})
require $F > 1$ to make
physical sense. We have checked numerically that $F$ as given by
equation (\ref{Fdef}) exceeds unity for all
disks in which the fluid velocity $v$ of the disk
is moderately smaller than $v_c$ for any value
of $\gamma$. When $v$ approaches the critical velocity $v_c$
(where the frame-dragging parameter $Q_0$ becomes very large),
numerical errors prevent us from confirming that prograde circular
orbits exist.  An analytic argument relieves our worries on this
score.

For a fixed value of $\gamma$, the solution sequence terminates
when $Q_0\rightarrow \infty$. Let us evaluate $F$ in this limit.
The equation of motion of the disk matter (\ref{eom1}) may be
written as
\begin{equation}
    1-n = \frac{1-v^2}{(Q_0v+1)(1+\gamma)}.
\end{equation}
Thus, when $Q_0\rightarrow \infty$, equation (\ref{Fdef}) with
the upper sign choice becomes
\begin{equation}
    F \rightarrow \frac{1}{2}\left[1 +\sqrt{1 +4v^2(1+\gamma)^2
    /(1-v^2)^2}\right],
\label{Fresult}
\end{equation}
which explicitly satisfies $F > 1$ for any values of $v >0$ and
$\gamma\ge 0$.  In the same limit and with the same upper sign choice,
the three-velocity of the test particle in equation (\ref{partvel})
is given by
\begin{equation}
    v_p 
    =\sqrt{ \frac{-1+\sqrt{1 +4v^2(1+\gamma)^2/(1-v^2)^2}}
    {1+\sqrt{1 +4v^2(1+\gamma)^2
    /(1-v^2)^2}}}.
\end{equation}
Notice that $v_p=v$ when $\gamma=0$. In
other words, the velocity of a test particle in a prograde
circular orbit equals the velocity of the disk matter when the
latter has zero pressure -- a satisfying consistency check of the
result.

We now wish to investigate whether circular orbits are
stable. Imagine perturbing the radial position $r$ of the test
particle about its equilibrium position $r_0$ by a small amount
$r_1$, keeping $\Et$ and $\lt$ fixed. To lowest non-vanishing
order upon expansion about $r_0$, equation (\ref{orbitODE})
becomes
\begin{equation}
    \left({dr_1\over d\tau}\right)^2 = - V''(r_0) r_1^2.
\end{equation}
Stability of the motion depends on the sign of $V''(r_0)$.  If
$V''(r_0) >0$, then circular orbits are stable, because there are
no perturbations -- maintaining the same specific energy and
angular momentum -- that will produce a real solution for $r_1$ in
the above equation.  In this case, of all orbits of a given
specific angular momentum $\lt$, the circular orbit has least
specific energy, therefore it is not possible to perturb the
circular orbit from its equilibrium without giving the particle
some additional specific energy, which will then cause it to
oscillate about the equilibrium radius $r_0$ with an ``epicyclic''
frequency $\sqrt{V''(r_0)}$.  The radial motion may be pictured as
rolling up and down the walls of a ``valley.'' On the other hand,
if $V''(r_0) < 0$, then circular orbits are unstable, because a
small perturbation of such an orbit -- even one that retains the
original specific energy and angular momentum -- will lead to
exponentially growing departures from the equilibrium radius
$r_0$.  In this case, of all orbits of a given specific angular
momentum $\lt$, the circular orbit has a (local) maximum of
specific energy, and the test particle becomes unstable by rolling
off a ``hill.''

In detail, after some algebra, we obtain
\begin{equation}
    V''(r_0) = (1-n)e^{N_0}r_0^{-2}
\left[ n\left(\sqrt{F}\pm Q_0\sqrt{F-1}\right)^2
    +(1-Q_0^2)(F-1)\right],
\end{equation}
where we have made use of the expressions for $\lt$ and $\Et$ from
equation (\ref{param-disk}). When $Q_0=0$ (and hence $v=0$), $1-n
= 1/(1+\gamma)$, $F = 1+\gamma$, and $V''(r_0) > 0$.  In this
limit, retrograde and prograde circular orbits are both stable.
In the limit $Q_0\rightarrow 1-$, $F\rightarrow 1/(1-n^2)$ for prograde
orbits, and $F(1-Q_0^2) \rightarrow (1+n)/(1-n)$ for retrograde orbits.
Thus, $V''(r_0)$ remains positive for both cases.
Retrograde and prograde circular
orbits are still stable forms of motion at the onset of
the disk's development of an ergoregion, although even the most
rapidly moving retrograde particle finds it difficult
to resist frame-dragging when $Q_0\rightarrow 1-$.  As $Q_0$
passes through unity and approaches
$\infty$, retrograde circular motion at velocities
less than the speed of light becomes impossible,
but the product $(1-n)Q_0\rightarrow
(1-v^2)/(1+\gamma)v$ remains finite and positive, so $V''(r_0) >
0$ for prograde circular orbits.
We have verified numerically that $V''(r_0)$ stays positive
between these various limits.
In summary, prograde circular orbits exist
and are stable for power-law isothermal disks
from the non-relativistic to the ultra-relativistic regime;
whereas retrograde circular orbits are possible and stable
only for disks that do not develop an ergoregion.

We may state the result in an alternative way that relates to
known results concerning circular orbits around black holes.  Of
all orbits of a given specific energy, the circular orbit in a
stable/unstable situation has a (local) maximum/minimum of
specific angular momentum. In the case of a Schwarzschild black
hole, we know that circular orbits that start off at great
distances from the event horizon, $r_0\gg r_{\rm Sch}$, are close
to the Newtonian limit, and therefore are stable. They remain
stable as long as the specific angular momentum of the circular
orbit continues to decrease with decreasing orbital radius (or
circumference).  There comes a point, $r_0=3r_{\rm Sch}$, when the
square of the ``epicyclic frequency'' $V''(r_0)$ changes sign, and
the specific angular momentum $\lt$ of the circular orbit has an
inflection point, and starts to increase for decreasing $r_0$.
This violation of ``Rayleigh's criterion'' signals a transition
from stable to unstable circular orbits.  Because of frame
dragging, the case of Kerr black holes is more complicated, but
can be similarly elucidated as we have done above for the disk
case.

In our power-law disks, every radius $r_0$ is similar to any other
radius, and spacetime is not flat even at infinity.
Thus, if a circular orbit is stable at any radius in a given
model, circular orbits at all radii are stable.  It is hard to
imagine how one could realistically construct a self-gravitating
disk of rotating material otherwise.

\section{Photon Orbits}
\label{Photon Orbits}
The case of photon orbits in the spacetime of relativistic SIDs
is also interesting.
Our self-similar metric equation (\ref{metric}) admits a homothetic Killing
$\xi$ satisfying
\begin{equation}
    {\cal L}_{\bf\xi} g_{\mu\nu} = 2\xi_{(\mu;\nu)} = 2
    g_{\mu\nu}
\end{equation}
In component form, it reads
\begin{equation}
    g_{\mu\nu ,\alpha} \xi^\alpha + g_{\mu\alpha}
    \xi^\alpha_{\phantom{\alpha},\nu} + g_{\alpha \nu}
    \xi^\alpha_{\phantom{\alpha},\mu} = 2g_{\mu\nu}
\end{equation}
The solution to this equation is
\begin{equation}
    \xi^\mu = [(1-n) t, 0,  r, 0]
\end{equation}
Associated with this vector is a conserved quantity $\Gamma =
\xi_\mu k^\mu$ for null geodesics $k^\mu = dx^\mu/d\lambda$, where
$\lambda$ is an affine parameter.  Indeed,
\begin{equation}
    {d\Gamma\over d\lambda} = \Gamma_{;\nu} k^\nu = (\xi_\mu
    k^\mu)_{;\nu} k^\nu = \xi_{(\mu;\nu)} k^\mu k^\nu = g_{\mu\nu}
    k^\mu k^\nu =-m^2,
\end{equation}
where we have used the geodesic equation $k^\mu_{\phantom{\mu};\nu}
k^\nu = 0$.  Therefore, for a massless particle, $m=0$, and
$\Gamma$ is a constant of motion.

In addition, we have the two ordinary Killing vectors associated
with the stationarity and axial symmetry of the spacetime:
\begin{equation}
    \xi^{(t)} = {\partial\over \partial t}, \qquad \xi^{(\phi)} =
{\partial \over \partial \phi}
\end{equation}
In total, we have the following three conserved quantities
\begin{equation}
    E = - k_t, \qquad l = k_\phi, \qquad \Gamma = (1-n)t k_t + r
    k_r \Rightarrow k_r = \frac{1}{r} \left[\Gamma + (1-n) t E\right]
\end{equation}

The null condition $k^\mu k_\mu = 0$ can be used to determine
$k_\theta$:
\begin{equation}
    k_t^2 g^{tt} + 2k_t k_\phi g^{t\phi} + k_\phi^2 g^{\phi\phi }
    + k_r^2 g^{rr} + k_\theta^2 g^{\theta\theta} = 0,
\end{equation}
which implies
$$   k_\theta = \pm \frac{1}{\sqrt{g^{\theta\theta}}}
    \left\{-E^2 g^{tt} + 2El g^{t\phi} - l^2 g^{\phi\phi} -
    \frac{1}{r^2} \left[\Gamma + (1-n) tE\right]^2 g^{rr}\right\}^{1/2}$$
\begin{equation}
    = \pm e^{Z/2}\left\{\left[Er^{1-n} e^{-N} - Q l e^{-P}\right]^2
    - l^2 e^{-2P} - \left[\Gamma + (1-n) Et\right]^2 e^{-Z}\right\}^{1/2}.
\end{equation}
Finally, the geodesic is described by
\begin{equation}
    k^t = E r^{-2n} e^{-N} - r^{-(1+n)} e^{-P} lQ,
\end{equation}
\begin{equation}
    k^\phi = r^{-(1+n)} e^{-P}E Q + r^{-2} l e^{N-2P} (1-Q^2),
\end{equation}
\begin{equation}
    k^r = e^{N-Z} r^{-1} \left[\Gamma + (1-n)Et\right],
\end{equation}
\begin{equation}
    k^\theta = \pm r^{-2} e^{N-Z/2} \left\{\left[ Er^{1-n}
    e^{-N} - Q l e^{-P}\right]^2- l^2
    e^{-2P} - \left[\Gamma + (1-n) Et\right]^2
    e^{ -Z}\right\}^{1/2}.
\end{equation}
Divide everything by $k^t$, and we have
\begin{equation}
    {d\phi\over dt} = \frac{Q + \alpha
(1-Q^2)}{1 - Q \alpha}r^{n-1} e^{N-P},
\end{equation}
\begin{equation}
    {d{\ln r}\over dt} =\frac{e^{2N-Z} r^{n-1}\beta}{1 - Q \alpha},
\end{equation}
\begin{equation}
    {d\Theta\over dt}  = \pm \frac{e^{N-Z/2}(1+n) r^{n-1}}{1-Q\alpha}
\left[(1-Q\alpha)^2- \alpha^2 - \beta^2e^{2N-Z}\right]^{1/2},
\end{equation}
where
\begin{equation}
    \alpha \equiv \frac{l}{E} r^{n-1} e^{N-P}, \qquad
\beta \equiv \left[\frac{\Gamma}{E}+(1-n)t\right]r^{n-1}.
\label{alphabeta}
\end{equation}

To make this system autonomous, we extend the space to
include $\alpha$ and $\beta$ as two of the variables.
The self-similarity of the problem makes it
convenient to define the reduced radius and time as
\begin{equation}
\zeta \equiv \ln r \qquad {\rm and} \qquad d\tau = r^{n-1}\, dt.
\end{equation}
We then have
\begin{equation}
    {d\phi\over d\tau} = \frac{Q+\alpha(1-Q^2)}{1-Q\alpha} e^{N-P},
\label{geodesicone}
\end{equation}
\begin{equation}
    {d\zeta\over d\tau} = \frac{e^{2N-Z} \beta}{1-Q\alpha},
\label{geodesictwo}
\end{equation}
\begin{equation}
    {d\Theta\over d\tau} = \pm (1+n) \frac{e^{N-Z/2}}{1-Q\alpha}
\left[(1-Q\alpha)^2 - \alpha^2 - \beta^2e^{2N-Z}\right]^{1/2},
\label{geodesicthree}
\end{equation}
\begin{equation}
    {d\beta\over d\tau} =(1-n)
\left[1-\frac{e^{2N-Z}\beta^2}{1-Q\alpha}\right],
\label{geodesicfour}
\end{equation}
\begin{equation}
    {d\alpha\over d\tau} = \alpha \left\{ (n-1)
\frac{e^{2N-Z}\beta}{1-Q\alpha} \pm (N'-P')(1+n)
\frac{e^{N-Z/2}}{1-Q\alpha}
\left[(1-Q\alpha)^2-\alpha^2 - \beta^2 e^{2N-Z}\right]^{1/2}\right\},
\label{geodesicfive}
\end{equation}
\begin{equation}
    {dt\over d\tau} = e^{\zeta(1-n)}
\label{geodesicsix}
\end{equation}

\subsection{Dynamics and Geometry of the Photon Trajectories}
We need to specify initial values for the six dependent variables
$(\phi, \zeta, \Theta, \beta, \alpha, t)$ and integrate forward in
$\tau$.  Without loss of generality, stationarity, axial symmetry,
and self-similarity imply that we may take $t=0$, $\phi=0$, and
$r=1$ (or $\zeta=0$) at $\tau=0$.  We shall also assume that all
photons begin by being emitted from the plane of the disk,
$\Theta=\Theta_0$.  Except for a set of measure zero (involving
photons travelling outward exactly along the rotation axis), we
shall find that this last assumption also results in no loss of
generality because even photons emitted by external sources
outside of the disk are soon bent to cross the disk plane.  This
behavior can be attributed to the existence of an adiabatic
invariant $J$ (see below), which places a constraint on the
trajectories different from those presented by the classical
integrals $l$, $E$, and $\Gamma$.  In any case, we are now left
with only two arbitrary specifications, the initial values
$\alpha=\alpha_*$ and $\beta=\beta_*$, from which equations
(\ref{alphabeta}) allow us to reconstruct the two constants of
motion,
\begin{equation}
l/E = \alpha_*, \qquad \Gamma/E = \beta_*.
\end{equation}

Only the combinations $l/E$ and
$\Gamma/E$ provide isolating integrals because the principle
of equivalence forbids photons having values of
$l$, $E$, and $\Gamma$ that differ by only a single
multiplicative factor from having fundamentally
different trajectories in spacetime.
By this method of counting, we see that
(self-similar) photon trajectories
are completely determined by well-behaved integrals
of motion, and thus chaos does not enter the problem.
Indeed, self-similarity and axial symmetry decouple
the three ODES for $\Theta$, $\alpha$, and $\beta$ from the rest of
the set of equations (\ref{geodesicone})--(\ref{geodesicsix}),
so that the other three variables $(\phi, \zeta, t)$ may be
computed by post-processing.  Although useful
for proving the absence of chaotic photon orbits in this problem,
the above discussion unfortunately lends little descriptive power
to the geometry of the photon orbits.  Instead, we adopt
the following approach for choosing the parameters
of the initial conditions.

Let us consider the photon orbit as seen by a locally nonrotating
observer (LNRO) whose basis one-forms are given by
\begin{equation}
    {\bf \omega}^{(t)} = r^n e^{N/2} dt,
\end{equation}
\begin{equation}
    {\bf \omega}^{(\phi)} = r e^{P-N/2}
(d\phi - r^{n-1} e^{N-P} Q dt),
\end{equation}
\begin{equation}
    {\bf \omega}^{(r)} = e^{(Z-N)/2} dr,
\end{equation}
\begin{equation}
    {\bf \omega}^{(\theta)} = r e^{(Z-N)/2}
d\theta.
\end{equation}
Projected onto this frame, $k^{(\mu)} = \mathbf{k} \cdot
{\bf \omega}^{(\mu)}$.  In component form,
\begin{equation}
    k^{(t)} = r^{-1}e^{N/2}\left[Er^{1-n}e^{-N}- e^{-P}lQ\right] = Er^{-n}
e^{-N/2} (1-Q\alpha),
\end{equation}
\begin{equation}
    k^{(\phi)} =r^{-1} l e^{N/2-P} = Er^{-n}e^{-N/2}\alpha,
\end{equation}
\begin{equation}
    k^{(r)} =e^{(N-Z)/2} r^{-1} [\Gamma + (1-n) Et] =
Ee^{(N-Z)/2}r^{-n} \beta,
\end{equation}
$$  k^{(\theta)} = \pm r^{-1} e^{N/2} \left\{\left[ E r^{1-n} e^{-N}
- Q l e^{-P}\right]^2- l^2e^{-2P} - \left[\Gamma + (1-n) Et\right]^2 e^{
b-Z}\right\}^{1/2}$$
\begin{equation}
     = \pm |E (1-Q\alpha)| r^{-n} e^{-N/2} \left[1 - \frac{\alpha^2
    + \beta^2 e^{2N-Z}}{(1-Q\alpha)^2}\right]^{1/2}.
\end{equation}
Notice that since $k^{(t)}$ is the energy measure by the LNRO, it
is always positive.  That means $E$ and $1-Q\alpha$ have the same
sign and we can safely omit the absolute sign in $k^{(\theta)}$.
Let $(\psi, \chi)$ be the direction of a photon trajectory seen by
this LNRO.  Then these angles are related to
equations (\ref{geodesicone})--(\ref{geodesicsix}) by
\begin{equation}
    \cos \psi = k^{(\theta)}/k^{(t)} = \pm
\left[ 1 - \frac{\alpha^2 + \beta^2
e^{2N-Z}}{(1-Q\alpha)^2}\right]^{1/2},
\end{equation}
\begin{equation}
    \tan \chi =k^{(\phi)}/k^{(r)} = e^{Z/2-N} \frac{\alpha}{\beta}.
\end{equation}
It is more intuitive to specify $\psi$ and $\chi$ as initial
conditions.  In particular, if we adopt the other choices
discussed at the beginning of this subsection, we have
at $\tau=0$:
\begin{equation}
    \phi = 0, \qquad
    \zeta = 0, \qquad
    \Theta = \Theta_0,
\end{equation}
\begin{equation}
    \beta = e^{Z/2-N} \frac{\sin\psi \cos \chi}{Q\sin \psi \sin
    \chi \pm 1} \equiv \beta_*, \qquad
    \alpha = \frac{\sin\psi \sin \chi}{Q\sin \psi \sin \chi \pm
    1} \equiv \alpha_*, \qquad
    t  = 0.
\end{equation}

\subsection{Sign Choices}
To solve the geodesic equations (\ref{geodesicone})--(\ref{geodesicsix})
numerically, we
need to be careful about the $\pm$ signs and the term in the
square root of
equation (\ref{geodesicthree}), which we call $\Lambda$:
\begin{equation}
    \Lambda \equiv (1-Q\alpha)^2 - \alpha^2 - \beta^2e^{2N-Z}.
\label{Lambda}
\end{equation}
In order for
$d\Theta/d\tau$ to be real, we need $d\Lambda/d\tau$ to vanish
whenever $\Lambda$ attains zero from positive values
(otherwise $\Lambda$ can become negative).  By straightforward
differentiation, it is easy to show
\begin{equation}
    -{1\over 2} {d\Lambda\over d\tau} =(Q-Q^2\alpha +\alpha)
{d\alpha\over d\tau} + \beta {d\beta\over d\tau} e^{2N-Z},
\end{equation}
where we have used that $d\Theta/d\tau$ = 0 when $\Lambda = 0$
to eliminate derivatives of functions of only $\Theta$.
Upon substitution of equations (\ref{geodesicfour}) and (\ref{geodesicfive}),
the last expression becomes
\begin{equation}
\left[Q+\alpha(1-Q^2)\right] \alpha (n-1) \frac{e^{2N-Z}
\beta}{1-Q\alpha} + \beta e^{2N-Z}(1-n)\left[ 1-\frac{e^{2N-Z}
\beta^2}{1-Q\alpha}\right].
\label{check}
\end{equation}
With $\Lambda=0$ in equation (\ref{Lambda}),
the term in the last square bracket equals
\begin{equation}
\left[\frac{(1-Q\alpha)Q\alpha
+\alpha^2}{1-Q\alpha}\right],
\end{equation}
and when this is substituted into equation (\ref{check}), we find
that the first and second terms algebraically cancel.
Thus, $d\Lambda/d\tau$ vanishes when $\Lambda = 0$, which
is the desired result.

Initially, we require that for $\psi < \pi/2$,
$d\Theta/d\tau$ is negative:
\begin{equation}
    {d\Theta\over d\tau} = -(1+n) e^{N-Z/2} \cos \psi
\end{equation}
Similarly, we want $d\zeta/d\tau$ positive whenever $-\pi/2
<\chi < \pi/2$.  This requires, after a little algebra,
\begin{equation}
    {d\zeta\over d\tau} = \pm e^{N-Z/2} \sin\psi \cos\chi > 0,
\end{equation}
i.e. we choose the $+$ sign in the initial conditions for $\alpha$ and
$\beta$.  We can then integrate
$\Theta$ forward in $\tau$ until $\Lambda = 0$, where we change
the sign of $d\Theta/d\tau$ (and correspondingly the sign in
$d\alpha/d\tau$).

\subsection{Results}
First we shall discuss the photon orbits in a spacetime without
the ergoregion ($Q_0<1$).  As an example, we choose the parameters
$n = 0.4$ and $\gamma = 0.5$ (or equivalently $v = 0.16$), and
consider photons launched in different directions as seen by the
LNRO. We assume that photons cross the disk plane without
absorption or scattering, an assumption that is more likely to
apply to massless low-energy neutrinos than real photons, for
which a disk with surface density $\sim 10^{23}$ g cm$^{-2}$ (if
we are talking about stellar-mass disks of sizes $\sim 1$ km) is
not likely to be optically thin. Our formal usage of the phrase
``photon orbits'' must henceforth be understood either as a
metamorphic rather than literal device, or as applicable only
while the photon is travelling above or below the disk plane.

\begin{figure}
\begin{center}
\includegraphics[height = .8 \linewidth, angle =
90]{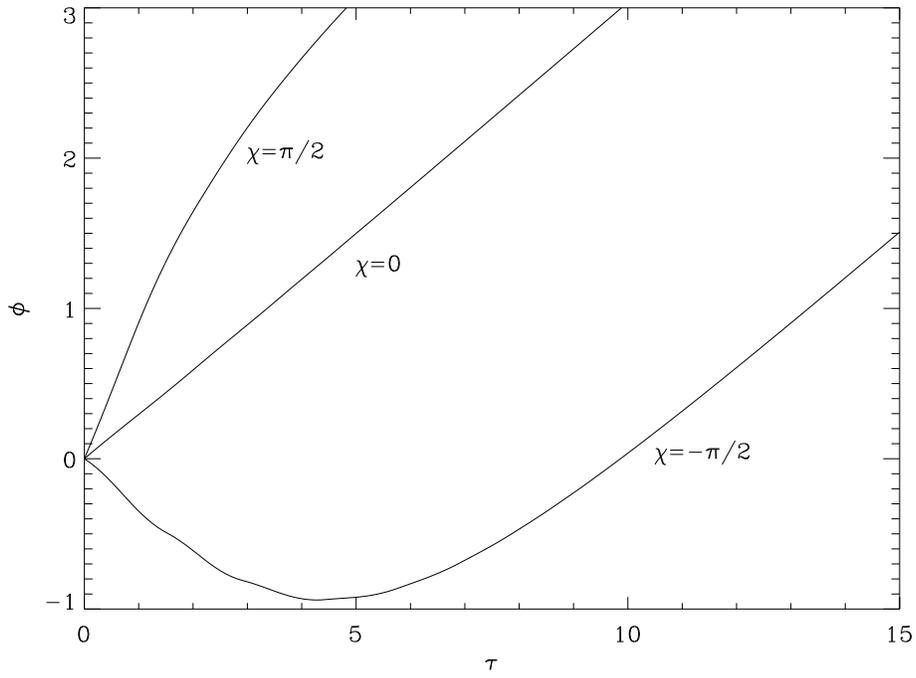}\\
(a) Azimuthal angle $\phi(\tau)$ for photons launched in different
directions\vskip .25in
\includegraphics[height = .8\linewidth, angle =
90]{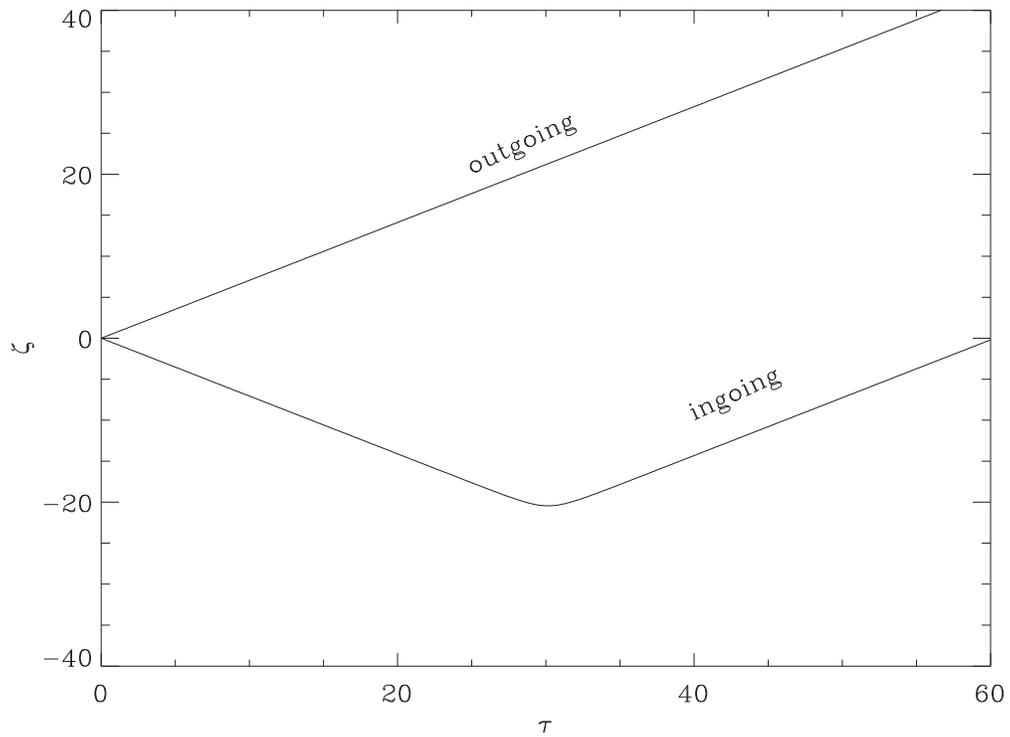}\\
(b) Logarithmic radius $\zeta(\tau)$ for ingoing and outgoing
photons\vskip .25in
\end{center}
\caption{Behavior of photon trajectories without the ergoregion.}
\label{noergo}
\end{figure}

We find that outgoing photons with $-\pi/2<\chi<\pi/2$
spiral out to $\zeta = \infty$, while ingoing photons with
$\chi>\pi/2$ or $\chi<-\pi/2$ reach a
minimum radius and then spiral out to $\zeta = \infty$ again.
Only
one photon launched along the disk with $\psi = \pi/2$) and
$\alpha = 0$ reaches the origin.  This was first discovered by
Lynden-Bell \& Pineault (1978b) for the cold
disk.  Photons launched in the retrograde direction,
reach a minimum coordinate angle $\phi_{min}$ and are then
dragged to go in the forward direction.  Figure \ref{noergo}
shows these orbits.

\begin{figure}[ht]
\begin{center}
\includegraphics[width = 4.0in, angle =90]{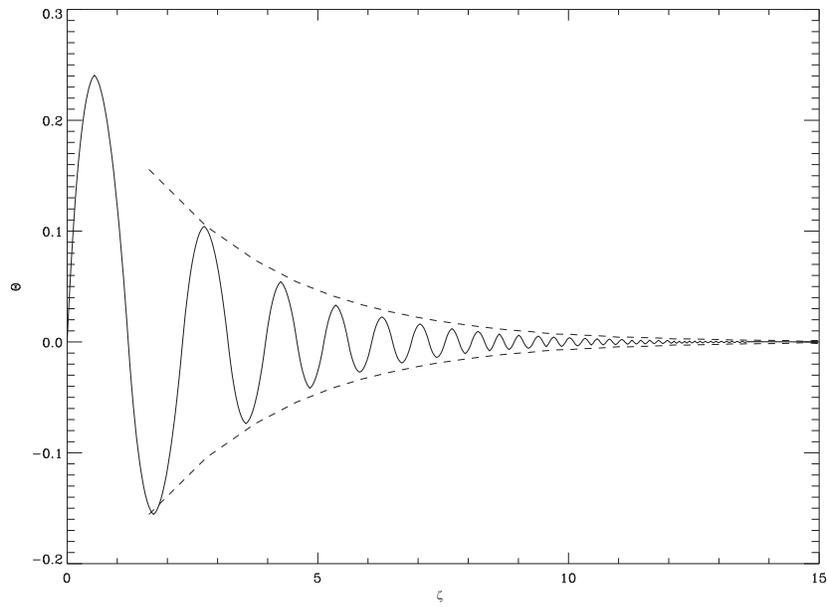}
\end{center}
\caption{Photon orbit showing $\Theta$ as a function of $\zeta$.
The turning point in
$\Theta$ decreases each time after the photon penetrates the
disk.  The solid line is from solving geodesic equation directly,
while the dotted line is the computed envelope from adiabatic
invariance.}
\label{theta_rho}
\end{figure}

We now discuss the out-of-plane behavior of the orbits. No matter
what is the initial condition, almost {\it all} orbits are focused
and eventually trapped by the disk. A (noninteracting) photon will
typically penetrate the disk many times. Each time it reaches a
turning point $\Theta_i$, where $i$ labels the number of
penetration (see Fig. \ref{theta_rho}).  The only trajectory that
can escape falling into the disk is one launched exactly along the
rotation axis. However, such a photon suffers an infinite redshift
as it propagates away from the disk, and an observer located at an
infinite distance above or below the disk cannot see it.  (This
result was first discovered for counter-rotating disks, and is
probably generic to self-similar configurations that contain an
infinite total mass.)

When interactions with the disk are ignored,
it appears that $\Theta_i$ decreases with each disk crossing $i$.
The result can be demonstrated to arise from adiabatic invariance.
The relevant invariant is easily computed conceptually.
The conjugate momentum to $\theta$ is $k_\theta$.  Hence we can define the
action integral,
\begin{equation}
    J \equiv \oint k_\theta d\theta
    =\frac{2}{1+n}|E|\int_{\Theta_{min}}^{\Theta_{max}} r^{1-n}
\left\{\left[1-Q\alpha\right]^2e^{Z-2N} -
\alpha^2e^{Z-2N} - \beta^2\right\}^{1/2}
d\Theta .
\end{equation}
Here we may treat $\alpha$ and $\beta$ as functions of $\Theta$
along the photon's trajectory during the current cycle.
The quantity $r(\Theta)$
varies slowly over this one cycle because it is a
monotonic function which does not oscillate.
Hence, we may approximate it by
its mean value.  To zeroth order, we may take
$r(\Theta)$ outside of the integral.  Thus
\begin{equation}
    J = |E|r^{1-n} \left[ {\cal J}(\Theta_{\rm max})-
{\cal J}(\Theta_{\rm min})\right].
\label{adiabinv}
\end{equation}
where
\begin{equation}
{\cal J}(\Theta_{\rm m}) \equiv \frac{2}{1+n}
\int_0^{\Theta_{\rm m}} \left\{\left[1-Q\alpha\right]^2
e^{Z-2N} - \alpha^2e^{Z-2N} - \beta^2\right\}^{1/2} d\Theta.
\end{equation}
Because the integrand is a known (numerical) function
of $\Theta$, we can tabulate the integral $\cal J $
as a function of its argument $\Theta_{\rm m}$
for each photon trajectory.
We can then invert the expression equation (\ref{adiabinv})
to recover $\zeta = \ln r$ as a function of
$\Theta_{\rm min}$, and $\Theta_{\rm max}$.
\begin{equation}
    \zeta = -\frac{1}{1-n}\ln \left[ {\cal J}(\Theta_{\rm max})
-{\cal J}(\Theta_{\rm min})\right] + C,
\end{equation}
where $C$ is the constant $\ln (J/E)$ divided by $1-n$.
The resulting curve is plotted as a dashed locus in Figure
\ref{theta_rho}.  The concordance between the dashed
curve and the actual envelope of the solid photon trajectory
is a measure of the goodness of the adiabatic invariant $J$.

It might be argued that the validity of the focusing effect
on photons seen in Figure \ref{theta_rho}
is compromised by the assumption that photons
continue in their original direction when they cross the disk
plane.  However, since all orbits, except for a set of measure
zero, always return to the disk plane, the qualitative
effect remains the same even if we were to allow photons
to interact with the disk matter. The absorption and re-emission
or the scattering of photons when they cross the disk would
result in a slow transfer of photons from the inner disk
to the outer disk because no photon can permanently escape
from the disk.  The focusing of photons toward the equatorial
plane is perhaps a generic feature of relativistic disks
(and not just peculiar to these self-similar configurations),
and may present an obstacle to some classes of beamed-jet models
for gamma-ray burst sources.

\begin{figure}[ht]
\begin{center}
\includegraphics[height =.75\textwidth, angle = 90]{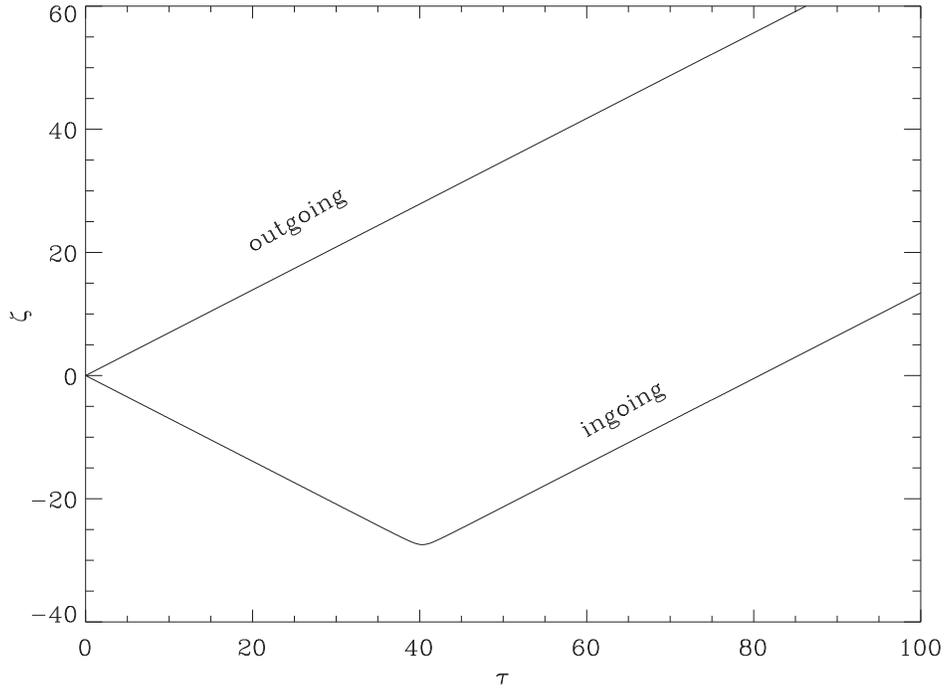}\\
(a) Logarithmic radius $\zeta(\tau)$ for ingoing and outgoing
photons\vskip .25in
\includegraphics[height =.75\textwidth, angle =
90]{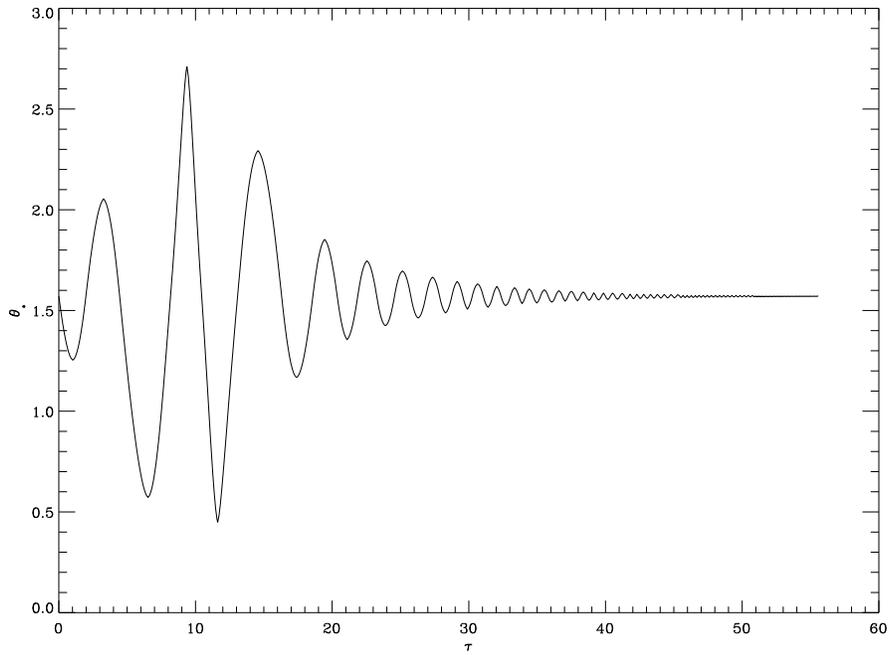}\\
(b) Normalized polar angle $\theta_*(\tau)$ for ingoing photons.
Here $\theta_*$ is normalized such that the disk is located at
$\pi/2$.\vskip .25in
\end{center}
\caption{Behavior of photon trajectories lauched in the forward
direction with an ergoregion.} \label{ergo_for}
\end{figure}

Photon orbits in a spacetime with $Q_0>1$ are more complicated
than those discussed so far.
As an example, let us take $n=0.75$, $v=0.32$ (and $\gamma = 0.5$).
All forwardly propagating photons ($\chi > 0$)
escape to $\zeta = \infty$.  The ones
launched outward ($\chi < \pi/2$) escape directly, with the
$\theta$ dependence mimicking the behavior of those in spacetime
without an ergoregion.  The ones
launched inward ($\chi > \pi/2$) reach a minimum
radius and then escape.  In the interior region, the photons spiral
toward the axis above and below the disk.  The turning
points in $\theta$ increase as the photons approach the minimum
radius.  Once past the
starting point
$\zeta = \ln r = 0$, we see the familiar $\theta$ behavior governed by
the adiabatic invariance described above.  Figure \ref{ergo_for} shows the
$\zeta$ and $\theta$ behavior of these trajectories.

\begin{figure}[ht]
\begin{center}
\includegraphics[height = \textwidth, angle = 90]{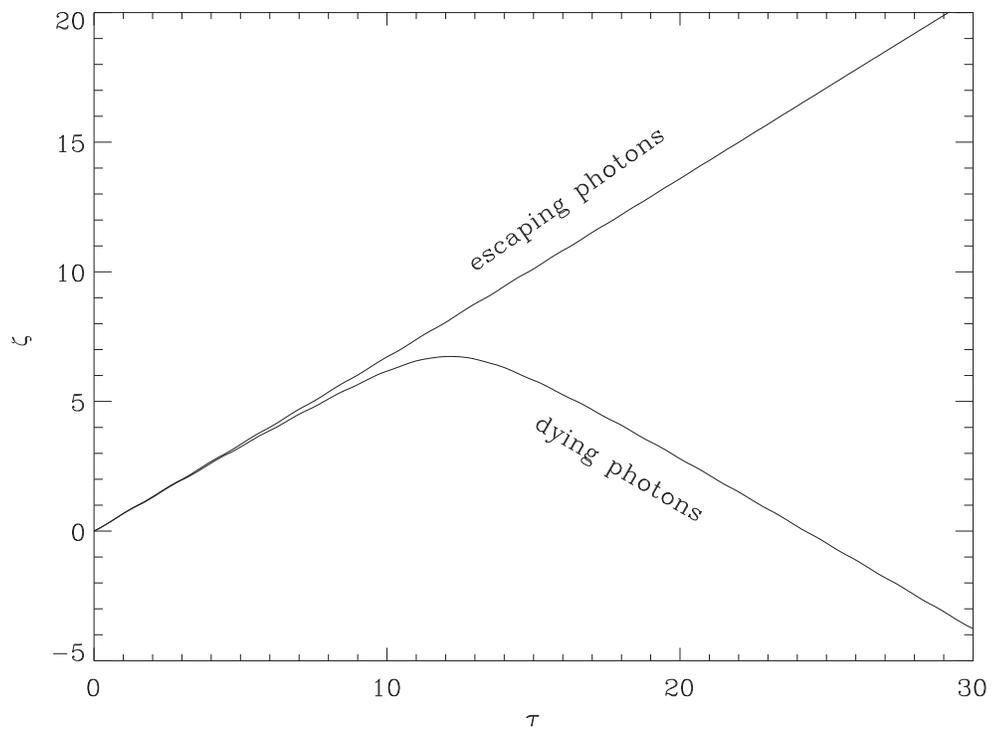}
\end{center}
\caption{Logarithmic radius $\zeta(\tau)$ for photons on either side of $E=0$ surface}
\label{ergo_back}
\end{figure}

\begin{figure}[ht]
\begin{center}
\includegraphics[height = \textwidth, angle = 90]{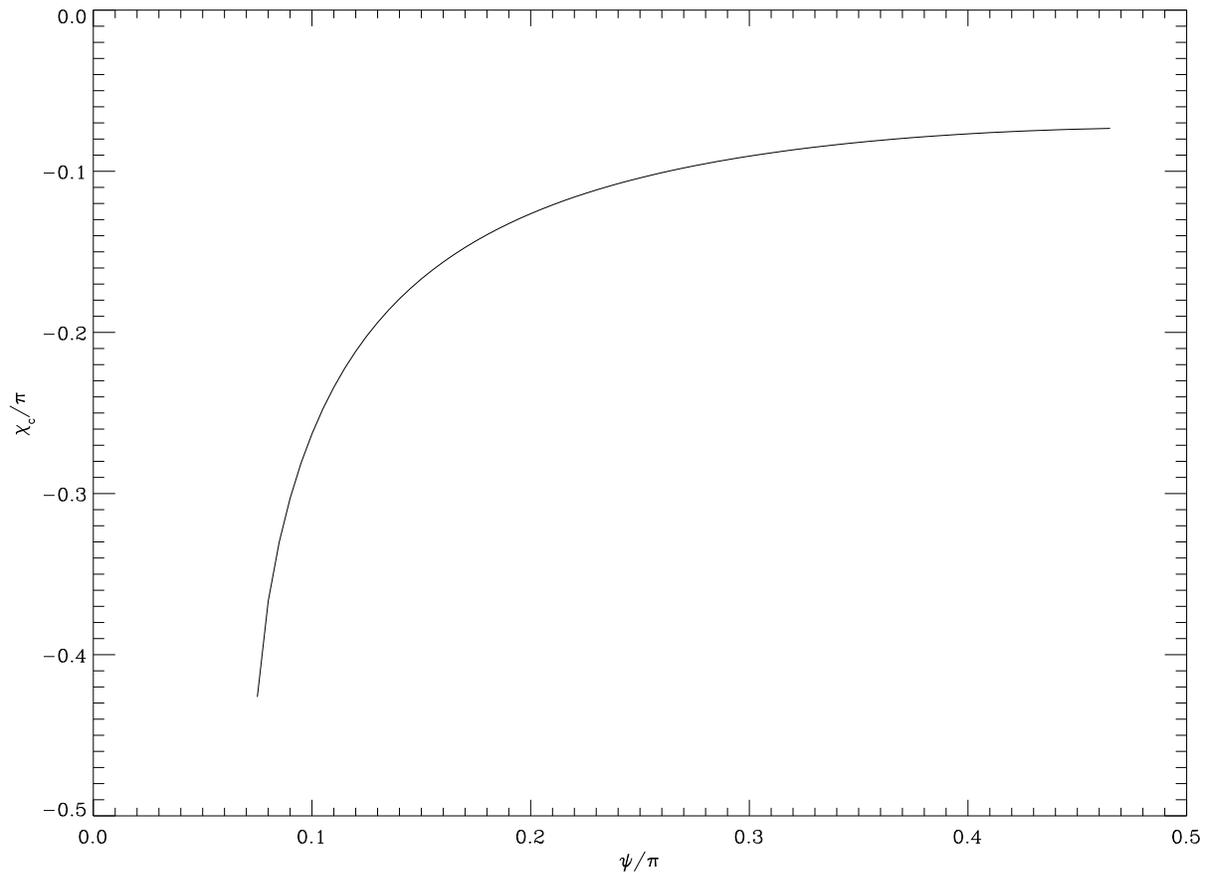}
\end{center}
\caption{Zero energy surface ($E=0$) represented by $\chi_c(\psi)$}
\label{ergo_sep}
\end{figure}

The backwards photons are divided into two classes, separately by the
surface $\chi_c(\psi)$.  Roughly speaking, this surface corresponds to
$E=0$ in the outgoing direction.  All ingoing photons fall towards the
origin directly.  For an outgoing photon, if $\chi > \chi_c$, then it
escapes to infinity.  On the other hand if $\chi < \chi_c$, the photon
reaches a maximum radius, and then falls to the origin (see Fig.
\ref{ergo_back}).  For the spacetime consider here, the $E=0$ surface
is plotted in Figure \ref{ergo_sep}.  Lynden-Bell \& Pineault
(1978b) gave an analytic expression for
$\chi_c(\pi/2) = \sin^{-1} (-1/Q_0)$.  For
this particular spacetime, $Q_0\approx 4.41$.  Therefore
$\chi_c(\pi/2) = -0.073$, which agrees with our numerical result.

\section{Conclusion}
\label{Conclusion}

We have solved by semi-analytic means the Einstein field equations
for axisymmetric, self-similar, relativistic
disks with ``flat'' rotation curves, including finite
levels of pressure support.  These spacetimes are not
asymptotically flat and cannot describe correctly the
behavior of isolated astrophysical objects when examined
at distances that are very large compared to their natural
gravitational radii.  Nevertheless, the solutions may yield
some insight into the near-field solutions of rapidly
rotating, compact objects.

As expected from first principles,
the solution space is parameterized by two dimensionless
numbers, $v$ and $\gamma$, that represent the disk rotation
speed and the square of the isothermal sound speed when
both are normalized appropriately by the speed of light $c$.  The
qualitative behavior of these disks resemble those found by
Lynden-Bell \& Pineault (1978b) for the cold disk.
This is encouraging because cold disks are known in their
Newtonian limits to be violently unstable to a wide variety
of spiral and barlike perturbations, and we cannot expect
their relativistic counterparts to behave much better.
A proper stability analysis of the disks discussed in
the current paper remains a task for the future.

Ergoregions develop for relatively low rotation velocities
in our disks and take the shape of (the outside of) a cone centered
around the axis.  As the rotational velocity increases, the
``ergo-cone'' closes up towards the axis.  For each $\gamma$, there is
a maximum velocity $v_c$ beyond which no equilibrium can exist due to
infinite frame dragging.  It should be noted that this
maximum velocity lies well below the special relativistic
limit of the speed of light.

We examined the behavior of test particles
with nonzero rest mass placed in circular orbits in the
plane of the disk.  We found that prograde circular orbits
exist and are stable for the full range of disk
models in this paper.  Retrograde circular orbits
are also stable when they exist, but cannot be maintained
against frame dragging by particle velocities less
than the speed of light when the disks develop ergoregions.

We also carried out a systematic study of planar and
non-planar photon orbits.  Most interestingly, we found that
all photon orbits are ultimately attracted toward the plane
of the disk because of the operation of a general
adiabatic invariant.  Although the formal result depends
on the disk being optically thin to the propagating photons
(an unlikely state of affairs), we gave physical arguments
why the generic effect may pose defocusing difficulties
for some classes of models of gamma-ray burst sources that
rely on beamed jets along the rotation axis of rapidly rotating
compact objects.  To be sure, the effect in realistic flattened
systems that do not have infinite mass and spatial extent
may be less dramatic than the one found here for relativistic
SIDs.   A lower bound on the effect might
be obtained by examining the analogous
properties of photon orbits in a Kerr geometry,
which in other respects mimics the spacetime analyzed
in the current paper.

It is our belief that the current investigation has just begun
to scratch the surface of a potentially very rich mine
for general relativity to explore.  As discussed in the
Introduction, the study of self-similar (Mestel) disks
in the Newtonian limit has uncovered rich veins
relating to the stability and collapse of such objects
that have illuminated astronomers' understanding of
real-world objects such as protoplanetary disks
and triaxial and spiral galaxies.  In addition to serving
as useful testbeds for numerical relativity codes,
the relativistic generalization of such studies
could shed light on topics such as the efficiency
of gravitational radiation and the possible generation
of naked singularities during gravitational collapse.

\acknowledgements

We thank Donald Lynden-Bell for useful conversations.  This
work has been supported by the National Science Foundation
by a Graduate Fellowship awarded to MJC and by grant
AST-9618491 awarded to FHS.

\newpage

\begin{figure}
\begin{center}
\includegraphics[angle = 90, width = 4.5in]{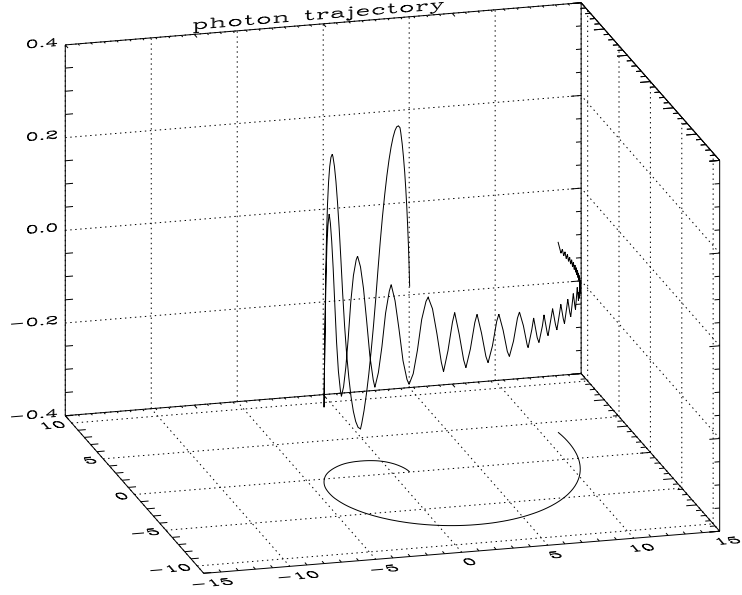}
\end{center}
\caption{A typical photon trajectory.  Here the distance is plotted on
a log scale, normalized such that the closest approach to the origin
is 1.  The nonoscillatory curve at the bottom is the projection
of the photon trajectory onto the equatorial plane of the disk.}
\end{figure}

\begin{figure}[ht]
\begin{center}
\includegraphics[width = 3.5in, angle = 90]{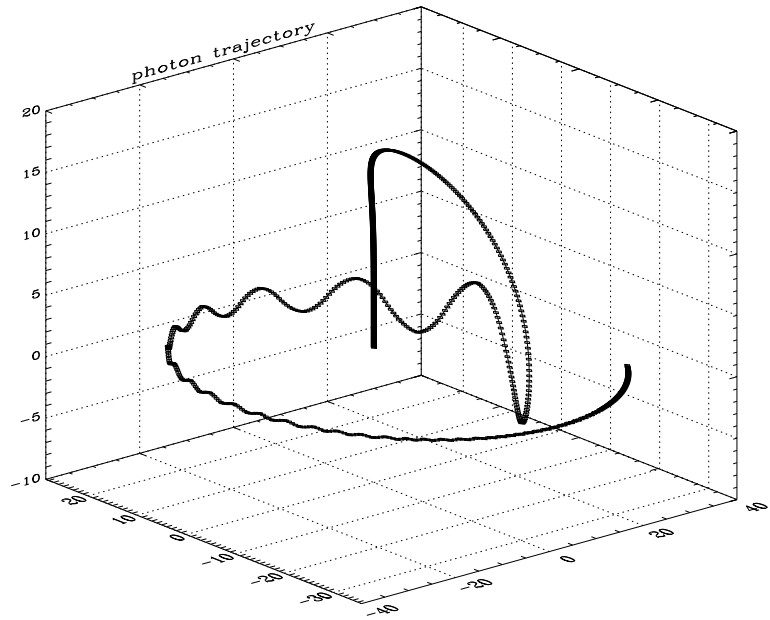}
\caption{A photon launched along $\Theta = 10^{-6}$
is being trapped by the disk.  Again, the distance is plotted on a
normalized log scale.}
\end{center}
\end{figure}


\newpage
\begin{deluxetable}{cc}
\tablecaption{Critical Velocity as a Function of Sound Speed Squared}
\label{vcritical}
\tablecolumns{2}
\tablewidth{0pt}
\tablehead{\colhead{$\gamma$}&\colhead{$v_c$}}
\startdata
    0.0&0.438\cr
    0.1&0.415\cr
    0.2&0.398\cr
    0.3&0.381\cr
    0.4&0.366\cr
    0.5&0.351\cr
    0.6&0.339\cr
    0.7&0.327\cr
    0.8&0.316\cr
    0.9&0.306\cr
    1.0&0.298\cr
\enddata
\end{deluxetable}

\begin{deluxetable}{cccccccccccc}
\tablecaption{$100\, \hat {\cal E}$ as a Function of $v/v_c$ and $\gamma$}
\label{surfdens}
\tablecolumns{12}
\tablewidth{0pt}
\tablehead{\colhead{\gvvc}&\colhead{0.0}&\colhead{0.1}&\colhead{0.2}&
\colhead{0.3}&\colhead{0.4}&\colhead{0.5}&\colhead{0.6}&\colhead{0.7}
&\colhead{0.8}&\colhead{0.9}&\colhead{1.0}}
\startdata
%          &0.0&0.1&0.2&0.3&0.4&0.5&0.6&0.7&0.8&0.9&1.0\cr
    0.0&0.0&0.0313&0.124&0.280&0.498&0.780&1.13&1.53&2.00&2.54&3.14\cr
    0.1&1.23&1.24&1.33&1.48&1.67&1.92&2.25&2.73&3.42&4.40&5.78\cr
    0.2&1.97&1.99&2.07&2.19&2.36&2.58&2.86&3.28&3.84&4.61&5.65\cr
    0.3&2.50&2.49&2.55&2.66&2.80&3.00&3.26&3.60&4.06&4.67&5.47\cr
    0.4&2.84&2.84&2.89&2.98&3.12&3.30&3.53&3.83&4.23&4.74&5.40\cr
    0.5&3.07&3.08&3.14&3.23&3.35&3.51&3.72&3.99&4.33&4.76&5.29\cr
    0.6&3.26&3.26&3.29&3.36&3.47&3.61&3.79&4.03&4.33&4.70&5.16\cr
    0.7&3.39&3.38&3.41&3.48&3.58&3.71&3.88&4.09&4.35&4.67&5.05\cr
    0.8&3.46&3.46&3.50&3.56&3.64&3.75&3.89&4.07&4.28&4.54&4.85\cr
    0.9&3.52&3.53&3.56&3.61&3.68&3.77&3.89&4.04&4.23&4.45&4.71\cr
    1.0&3.56&3.57&3.60&3.64&3.71&3.80&3.91&4.05&4.22&4.42&4.66\cr
\enddata
\end{deluxetable}

\end{document}